\documentstyle[aps]{revtex}

\begin{document}

\title{Theory of Electronic Ferroelectricity}

\author{T. Portengen \footnote{Present address: Clarendon
Laboratory, University of Oxford, Parks Road, Oxford OX1
3PU, United Kingdom}, Th. \"{O}streich \footnote{Present
address: Institut f\"{u}r Theoretische Physik, Universit\"{a}t
G\"{o}ttingen, Bunsenstrasse 9, D-37073 G\"{o}ttingen,
Germany}, and L. J. Sham}

\address{Department of Physics, University of California
San Diego, La Jolla, California 92093-0319}

\date{28 May 1996}

\maketitle

\begin{abstract}
We present a theory of the linear and nonlinear optical
characteristics of the insulating phase of the Falicov-Kimball
model within the self-consistent mean-field approximation.
The Coulomb attraction between the itinerant $d$\/-electrons
and the localized $f$\/-holes gives rise to a built-in coherence
between the $d$\/- and $f$\/-states, which breaks the inversion
symmetry of the underlying crystal, leading to: (1) electronic
ferroelectricity, (2) ferroelectric resonance, and (3) a
nonvanishing susceptibility for second-harmonic generation.
As experimental tests of such a built-in coherence in mixed-valent
compounds we propose measurements of the static dielectric
constant, the microwave absorption spectrum, and the dynamic
second-order susceptibility.
\end{abstract}


\section{Introduction}
In this paper, we present a theory of ferroelectricity originating
from an electronic phase transition, in contrast to the conventional
displacive ferroelectricity due to a lattice distortion~\cite{Kittel}.
The electronic ferroelectricity occurs in a strongly correlated electron
system, namely, the insulating phase of the Falicov-Kimball model.

The Falicov-Kimball (FK) model was introduced originally~\cite{Falicov}
as a simple model to explain the metal-insulator transitions observed
in certain transition-metal and rare-earth compounds. The model has
since been applied extensively to the mixed-valent compounds and
heavy-fermion materials. The FK model introduces two types of
electrons: itinerant $d$\/-electrons and localized $f$\/-electrons.
The valence transition is driven by the on-site Coulomb repulsion
between the $d$\/- and $f$\/-electrons. A $d$\/-$f$ hybridization
term may or may not be added to the model. The theoretical
solutions for the ground state of the FK model can be divided
into two classes. On the one hand, solutions with $f$\/-occupation
as a good quantum number~\cite{Freericks,Farkasovsky} do not have
a built-in coherence between $d$\/-electrons and $f$\/-holes. On
the other hand, solutions such as the self-consistent mean-field
(SCMF) solution~\cite{Leder} and the electronic polaron~\cite{Liu}
do have a built-in coherence between $d$\/-electrons and $f$\/-holes.

The built-in coherence of the SCMF solution breaks the inversion
symmetry of the FK Hamiltonian in the following way. The many-electron
eigenstates of the Hamiltonian can be classified into even and odd
parity states. The SCMF ground states with even and odd parities are
degenerate. A linear combination of the even and odd parity states
forms the appropriate ground state for the limit of a vanishing
electric field. We shall argue that such degenerate ground states
can exist for solutions with $f$\/-occupation as a good quantum
number. Thus, the inversion symmetry breaking is not limited to
the SCMF solution.

The primary purpose of this paper~\cite{Portengen} is to give a
detailed account of the linear and nonlinear optical characteristics
of the SCMF solution. The inversion-symmetry-broken ground state
possesses the following distinctive properties:
(1) electronic ferroelectricity, (2) ferroelectric resonance, and
(3) a nonvanishing susceptibility for second-harmonic generation.
Solutions of the model without built-in coherence do not have these
properties. As experimental tests to distinguish between the two classes
of solutions we propose measurements of the static dielectric constant,
the microwave absorption spectrum, and the dynamic second-order
susceptibility of a mixed-valent compound, for example ${\rm SmB}_{6}$.

In recent years, Four-Wave-Mixing (FWM) spectroscopy has
emerged as a powerful tool for studying coherence in optically
pumped semiconductor systems~\cite{Leo,Weiss,Kim}.
In a three-beam FWM experiment, two incoming beams of wavevectors
${\bf k}_{1}$ and ${\bf k}_{2}$ set up a transient polarization
grating. A third incoming beam of wavevector ${\bf k}_{3}$
diffracts off this grating to produce an outgoing signal in the
direction ${\bf k}_{4}={\bf k}_{3}+{\bf k}_{2}-{\bf k}_{1}$.
Being a third-order process, FWM is allowed in media with
or without inversion symmetry. We pose the question: what happens
if the system being probed already has a polarization built into
it by nature? An example of such a system is the SCMF solution
of the FK model resulting in the Bose-Einstein condensation
of $d$\/-$f$ excitons.

As shown below, the built-in polarization leads to a nonlinear
optical response to {\em second} order in the external field.
The mixed-valent system has a nonvanishing susceptibility
$\chi^{(2)}(2\omega,\omega,\omega)$ for second-harmonic
generation. The built-in polarization replaces one of the incoming
beams of the three-beam FWM experiment. In crystals with
inversion symmetry, second-harmonic generation is forbidden in
the electric-dipole approximation. In the mixed-valent system
the built-in polarization breaks the inversion symmetry, allowing
second-harmonic generation to take place. We present a calculation
of the second-harmonic susceptibility of a model mixed-valent system
within the SCMF approximation. The second-harmonic susceptibility is
directly proportional to the built-in coherence $\Delta$, showing
that second-harmonic generation can be used as a test of $d$\/-$f$
exciton condensation in mixed-valent compounds.

The existence of a built-in polarization in the ground state
also means that, according to the SCMF theory, mixed-valent
compounds are ferroelectric. Whereas in commonly known ferroelectrics
the built-in polarization is due to the relative displacement of
positive and negative ions, the ferroelectricity in mixed-valent
compounds is of purely electronic origin. Apart from possible
Jahn-Teller distortions as a result of the electronic polarization,
 the valence transition does not involve a
change in the crystal structure. As shown below, the valence
transition is accompanied by a divergence of the static dielectric
constant at the critical value of the $f$\/-level energy. The
divergence of the static dielectric constant should be observable
in real mixed-valent compounds, for example by varying the
external pressure or temperature.

The continuous symmetry associated with the phase of $\Delta$ leads
to a Goldstone mode in the excitation spectrum of the mixed-valent
compound. In the pseudo-spin picture, the Goldstone mode corresponds
to a uniform precession of the pseudo-spins around the $z$\/-axis.
Ferroelectric resonance occurs when an ac electric field is applied whose
frequency coincides with that of the Goldstone mode. This phenomenon is
the electric analogue of magnetic resonance in ferromagnetic insulators.
The ferroelectric resonance frequency is proportional to the square root
of the effective bias field, and may depend on the sample shape, domain
structure, and crystal fields. In real mixed-valent compounds, ferroelectric
resonance should occur in the microwave regime.

The remainder of the paper is organized as follows. In
section~\ref{sec:MODEL} we discuss the SCMF solution for the ground state
of the FK model. We show that the SCMF ground state has a built-in
polarization which breaks the inversion symmetry of the FK Hamiltonian.
We calculate the static dielectric constant of a model mixed-valent system
using mean-field theory. In section~\ref{sec:CHIONE} we calculate the linear
susceptibility of the model system using the pseudo-spin formalism. We obtain
an analytical expression for the linear susceptibility in a uniform static
electric field. We determine the ferroelectric resonance frequency
and analyze the shape of the infrared absorption spectrum. In
section~\ref{sec:CHITWO}
we compute the second-harmonic susceptibility of the model system. In
section~\ref{sec:EXPERIMENT} we compare the results of the model calculation
to experimental data and propose experimental tests of coherence in real
mixed-valent compounds. In section~\ref{sec:DISCUSSION} we discuss the
possibility of ferroelectricity, ferroelectric resonance, and second-harmonic
generation for solutions of the FK model {\em other} than the SCMF solution.
The main results are summarized in section~\ref{sec:SUMMARY}.

\section{Model}
\label{sec:MODEL}
Ignoring the electron spin, the FK Hamiltonian is
\begin{eqnarray}
\label{eq:FKH}
H & = &
       \sum_{\bf k} \varepsilon_{\bf k} d^{\dagger}_{\bf k} d_{\bf k} +
E'_{f} \sum_{\bf k}  f^{\dagger}_{\bf k} f_{\bf k} +
       \sum_{\bf k} V_{\bf k} d^{\dagger}_{\bf k} f_{\bf k} +
                    {\rm h.c.}
\nonumber \\ & &
  + \frac{U}{N} \sum_{\bf k,k',q} d^{\dagger}_{\bf k +q} d_{\bf k }
                                  f^{\dagger}_{\bf k'-q} f_{\bf k'}.
\end{eqnarray}
Here $d^{\dagger}_{\bf k}$ creates a {\em d}\/-electron of momentum
${\bf k}$ and energy $\varepsilon_{\bf k}$, and $f^{\dagger}_{\bf k}$
creates an {\em f}\/-electron of momentum ${\bf k}$ and energy $E'_{f}$.
The parameter $U$ is the Coulomb repulsion between the {\em d}\/- and
{\em f}\/-electrons, $V_{\bf k}$ is the hybridization energy,
and $N$ is the number of sites. We assume that the $d$\/-band and the
$f$\/-level are derived from $d$\/- and $f$\/-orbitals on the same
site. We have chosen the $d$\/- and $f$\/-orbitals to yield a finite
dipole moment between them in the $z$\/-direction. For simplicity, we
consider a model system with a $d$\/-band of bandwidth $W$ and constant
density of states $\rho_{0}=1/(2W)$.

\subsection{SCMF solution}

The SCMF solution is analogous to the BCS theory of superconductivity
except that the pairing now occurs between a $d$\/-electron of
momentum ${\bf k}$ and an $f$\/-hole of momentum $-{\bf k}$
(Ref.~\onlinecite{Frohlich}). Whereas a Cooper pair carries charge,
an electron-hole pair is neutral. The SCMF solution therefore describes
an insulator, rather than a superconductor. Pairing between electrons
and holes may also occur in a semiconductor placed in an intense
coherent laser field~\cite{Schmitt}. In that case, the pairing is
externally enforced by the pump field. The mean-field decoupling of
the Hamiltonian~(\ref{eq:FKH}) yields the effective one-particle
Hamiltonian
\begin{eqnarray}
H & = & \sum_{\bf k}
   (\varepsilon_{\bf k}+U n_{f}) d^{\dagger}_{\bf k}d_{\bf k} +
   (E'_{f}+U n_{d}) \sum_{\bf k} f^{\dagger}_{\bf k}f_{\bf k}
\nonumber \\ & &
  + \sum_{\bf k} (V_{\bf k}-\Delta) d^{\dagger}_{\bf k}f_{\bf k}
  + {\rm h.c.} - N U n_{f} n_{d} + \frac{N}{U} |\Delta|^{2} ,
\end{eqnarray}
where $\Delta = \frac{U}{N} \sum_{\bf k} \langle\psi| f^{\dagger}_{\bf k}
d_{\bf k} |\psi\rangle$ is the gap parameter, and $n_{d} = \frac{1}{N}
\sum_{\bf k} \langle\psi| d^{\dagger}_{\bf k}d_{\bf k}|\psi\rangle$
and $n_{f} = \frac{1}{N} \sum_{\bf k} \langle\psi| f^{\dagger}_{\bf k}
f_{\bf k}|\psi\rangle = 1-n_{d}$ are the $d$\/-band and $f$\/-level
occupancies respectively. The SCMF ground state is
\begin{math}
|\psi\rangle = \prod_{\bf k}
( u_{\bf k} +
  v_{\bf k} d^{\dagger}_{\bf k}f_{\bf k} ) |0\rangle ,
\end{math}
where $|0\rangle$ is the state with no {\em f}\/-holes (the normal
state), and $u_{\bf k} = \cos \frac{1}{2}\theta_{\bf k}$
           ($v_{\bf k} = \sin \frac{1}{2}\theta_{\bf k}$) is the
probability amplitude for the pair state $({\bf k},-{\bf k})$ to
be occupied (unoccupied). The gap parameter $\Delta$ and the
$f$\/-level occupancy $n_{f}$ must be determined self-consistently
from
\begin{eqnarray}
\label{eq:GAP}
\Delta & = & \frac{U}{N} \sum_{\bf k} \frac{\Delta-V_{\bf k}}{2 E_{\bf k}} , \\
\label{eq:FOCC}
n_{f} & = & \frac{1}{2N} \sum_{\bf k} \left( 1 +
\frac{\xi_{\bf k}}{E_{\bf k}} \right) .
\end{eqnarray}
Here $E_{\bf k}=\sqrt{\xi^{2}_{\bf k} + |\Delta-V_{\bf k}|^{2}}$
is the quasiparticle excitation energy,
with $\xi_{\bf k}=\frac{1}{2}(\varepsilon_{\bf k}-E_{f})$, where
$E_{f} = E'_{f}+U n$ is the renormalized $f$\/-level energy, and
$n=n_{d}-n_{f}$ is the inversion. Since neither $E_{f}$ nor $E'_{f}$
are known from first principles, we shall treat $E_{f}$ as the materials
parameter. Eqs.~(\ref{eq:GAP}) and (\ref{eq:FOCC}) are Eqs.~(11) and (10) of
Ref.~\onlinecite{Leder} at temperature $T=0$, with a ${\bf k}$-dependent
hybridization. If the crystal has inversion symmetry, the hybridization
must satisfy $V_{-\bf k} = -V_{\bf k}$. The assumption of a
${\bf k}$-independent hybridization in, among others,
Ref.~\onlinecite{Leder}, is therefore incorrect. If, instead, we assume
nearest-neighbour hybridization, we find that $V_{\bf k}$ is odd
in ${\bf k}$ and purely imaginary. As can be seen from Eq.~(\ref{eq:GAP}),
the imaginary part of $\Delta$ then vanishes due to the cancellation of terms
with $\pm {\bf k}$. The real part of $\Delta$ is given by the BCS gap
equation
\begin{equation}
\label{eq:BCSGAP}
\Delta = \frac{U}{N} \sum_{\bf k} \frac{\Delta}{2 E_{\bf k}} ,
\end{equation}
where $E_{\bf k}=\sqrt{\xi_{\bf k}^2+|V_{\bf k}|^{2}+\Delta^{2}}$.
Calculation shows that a sufficiently strong hybridization can
destroy the gap. In the following we consider the limit where
$V_{\bf k}$ is negligible compared to $U$.

The solution of Eqs.~(\ref{eq:FOCC}) and (\ref{eq:BCSGAP}) for our model
system is shown in Fig.~\ref{fig:DELTA}.
The figure shows the gap parameter $\Delta$ and the $f$\/-level
occupancy $n_{f}$ as a function of the $f$\/-level energy $E_{f}$.
The gap parameter $\Delta$ is the order parameter of the valence
transition. When the $f$\/-level is far below the bottom of the
$d$\/-band, the system is in the normal state with no $f$\/-holes,
and $\Delta=0$. As the $f$\/-level is moved upward
past the critical value $E_{f}=-E_{c}$, where $E_{c}=W\coth(W/U)$
(in a real material this is achieved by applying pressure or by alloying),
$\Delta$ becomes nonzero and the system undergoes a valence transition.
In the mixed-valent state, the $f$\/-level occupancy $n_{f}$
lies between 0 and 1. The gap parameter reaches the maximum value
$\Delta(0)=W/[2\sinh(W/U)]$ when the $f$\/-level lies at the center of
the $d$\/-band ($E_{f}=0$). This is the half-filling point $n_{f}(0)=1/2$.
For still higher $E_{f}$, the $f$\/-level gradually empties out into
the $d$\/-band. The solutions transform according to
$\Delta(E_{f})=\Delta(-E_{f})$ and $n_{f}(E_{f})=1-n(-E_{f})$. At
$E_{f}=E_{c}$, the system returns to a normal state with
no $f$\/-electrons. Since $n_{f}(E_{f})$ has no discontinuities,
the valence transition is continuous for all values of the Coulomb
repulsion $U$.

\subsection{Inversion symmetry breaking}

The key feature of the SCMF ground state $|\psi\rangle$ is that it
breaks the inversion symmetry of the FK Hamiltonian. In general,
symmetry breaking occurs when the ground state has a lower symmetry
than the Hamiltonian. As a well-known example in another area, the
Ising model $H_{\rm Ising}=-J\sum_{<ij>}S_{i}^{z}S_{j}^{z}$ is
invariant under $S_{i}^{z}\rightarrow-S_{i}^{z}$, all $i$. Yet the
ground state is either one of the broken-symmetry states with
built-in magnetization $\pm M_{z}$. The sign of the magnetization
is selected by applying an infinitesimal bias field $H_{z}$, and
then choosing the lower-energy state.

In the case of the SCMF solution of the FK model, the inversion
symmetry in a $d$\/-$f$ site is spontaneously broken due to the
pairing of $d$\/-states of even parity with $f$\/-states of odd parity.
Applying the inversion $\hat{J}$ to $|\psi\rangle$ gives the state
\begin{equation}
\hat{J}|\psi\rangle = \prod_{\bf k} (-u^{*}_{\bf k}+v^{*}_{\bf k}
d^{\dagger}_{\bf k} f_{\bf k}) |\psi\rangle ,
\end{equation}
which is linearly independent of $|\psi\rangle$.  The form of the
inversion image follows from the even parity of the $d$\/-orbital and
the odd parity of the $f$\/-orbital and $u_{-\bf k} = u^{*}_{\bf k}$
and $v_{-\bf k} = v^{*}_{\bf k}$. The states
$|\psi\rangle$ and $\hat{J}|\psi\rangle$ have built-in polarizations
$\langle \psi|\hat{\bf P}|\psi\rangle = {\bf P}^{(0)}$ and
$\langle \psi|\hat{J}\hat {\bf P}\hat{J}|\psi\rangle = -{\bf P}^{(0)}$,
where
\begin{equation}
\hat{\bf P} = \frac{\mbox{\boldmath $\mu$}}{\Omega}
\sum_{\bf k} d^{\dagger}_{\bf k} f_{\bf k} + {\rm h.c.}
\end{equation}
is the polarization operator. \mbox{\boldmath $\mu$} is the interband
dipole matrix element, which for simplicity we take to be independent
of ${\bf k}$, and $\Omega$ is the volume. One can show that in the
thermodynamic limit $N \rightarrow \infty$, $|\psi\rangle$ and
$\hat{J}|\psi\rangle$ are orthogonal and uncoupled by $H$. The proof
relies on the fact that the infinite product
$\prod_{\bf k}(|v_{\bf k}|^{2}-|u_{\bf k}|^{2})$ is zero, since
$|(|v_{\bf k}|^{2}-|u_{\bf k}|^{2})|<1$ for almost all ${\bf k}$.

Because $H$ is invariant under inversion, $|\psi\rangle$ and
$\hat{J}|\psi\rangle$ are degenerate. As in the case of the
Ising model, the correct ground state is selected by lifting the
degeneracy with a bias field ${\bf E}$, and then choosing
the lower-energy state. This yields the state $|\psi\rangle$ with
built-in polarization in the direction of ${\bf E}$. We call
this direction the $z$\/-direction. (Without crystal-field terms,
the $z$\/-direction has no definite orientation with respect to the
crystal axes.) Since $\mu_{z}$ is real, $P^{(0)}_{z} = N \mu_{z}(\Delta
+ \Delta^{*})/(\Omega U)$, where $\Delta$ is the built-in coherence.
The built-in polarization vanishes in the normal state where
$\Delta = 0$.

\subsection{Electronic Ferroelectricity}

In commonly known ferroelectrics such as BaTiO$_{3}$, the ferroelectric
transition involves a change in the crystal structure. In the
ferroelectric phase, the positive ions are displaced relative to the
negative ions, leading to a permanent electric dipole moment. The
displacive ferroelectric transition occurs when the transverse
optical (TO) phonon frequency vanishes at some point in the Brillouin
zone. In an electronic ferroelectric, the ferroelectric transition
involves a change in the electronic structure rather than the crystal
structure. (Here we neglect the electron-phonon coupling, which may
cause a lattice distortion as a secondary effect of the transition.)
Instead of a vanishing of the TO phonon frequency, the $d$\/-$f$
exciton energy goes to zero at the critical value of the $f$\/-level
energy. The built-in polarization of an electronic ferroelectric is
of the order of $10$~$\mu$C/cm$^{2}$, comparable to the built-in
polarization of perovskites~\cite{Kittel}.

Since the built-in polarization is continuous at $E_{f}=\pm E_{c}$
(see inset of Fig.~\ref{fig:EPS}), the valence transition is a
second-order ferroelectric transition. In general, second-order
ferroelectric transitions are accompanied by a divergence of the
static dielectric constant in the direction of the spontaneous
polarization. For a temperature-driven transition, the dielectric
constant diverges as $(T-T_{c})^{-\gamma}$ above $T_{c}$, and as
$(T_{c}-T)^{-\gamma'}$ below. Almost all known ferroelectrics have
$\gamma=1$ (Curie-Weiss law). Observed values of $\gamma'$ range
from $1/2$ in TGS to $1/6$ in SbSI~\onlinecite{Fatuzzo}.

Here we calculate the static dielectric constant of the model
mixed-valent system at temperature $T=0$, using mean-field theory.
The dielectric constant in the $z$\/-direction is given by
$\epsilon_{zz} = 1+4\pi \chi^{(1)}_{zz}$, where $\chi^{(1)}_{zz} =
\lim_{E_{z}\rightarrow 0}\,\partial P^{(0)}_{z}/ \partial E_{z}$
is the static susceptibility. For $\Delta$ real, the
polarization is $P^{(0)}_{z} = 2N \mu_{z}\Delta/(\Omega U)$, so that
$\chi^{(1)}_{zz} = [2 N\mu_{z}/(\Omega U)]\,\lim_{E_{z}\rightarrow 0}
\partial \Delta/\partial E_{z}$. The bias field $E_{z}$ leads to
an additional term $-\mu_{z} E_{z} \sum_{\bf k}d^{\dagger}_{\bf k}
f_{\bf k} + {\rm h.c.}$ in the effective one-particle Hamiltonian.
The self-consistency equation for $\Delta$ then becomes
\begin{equation}
\label{eq:GAPEZ}
\Delta = \frac{U}{N} \sum_{\bf k}
\frac{\Delta + \mu_{z}E_{z}}{2E_{\bf k}} ,
\end{equation}
where $E_{\bf k}=\sqrt{\xi_{\bf k}^{2}+(\Delta+\mu_{z}E_{z})^{2}}$.
The susceptibility is obtained by implicit differentiation
of Eqs.~(\ref{eq:FOCC}) and (\ref{eq:GAPEZ}) with respect to $E_{z}$.
The susceptibility of the normal state ($\Delta=0$) is
\begin{equation}
\label{eq:CHI_N}
\chi^{(1)}_{zz} = -\frac{2 N \mu_{z}^{2}}{\Omega}\,
\frac{{\rm arccoth}(|E_{f}|/W)}{W - U {\rm arccoth}(|E_{f}|/W)} ,
\end{equation}
and the susceptibility of the mixed-valent state ($\Delta>0$) is
\begin{equation}
\label{eq:CHI_MV}
\chi^{(1)}_{zz} = -\frac{2N\mu_{z}^{2}}{\Omega U}
\left(1+\frac{1+4\Delta^{2}A(0)}{
4\Delta^{2}A(0)[1+4\Delta^{2}A(0)]+4\Delta^{2}B^{2}(0)}\right) .
\end{equation}
Here $A(0)$ and $B(0)$ are given by Eqs.~(\ref{eq:A0}) and
(\ref{eq:B0}) of the Appendix.
Fig.~\ref{fig:EPS} shows the dielectric constant of the model
system as a function of the $f$\/-level energy. The dielectric
constant diverges at $|E_{f}|=E_{c}$. From Eq.~(\ref{eq:CHI_N})
we find $\chi^{(1)}_{zz} \propto (|E_{f}|-E_{c})^{-1}$ as $|E_{f}|$
approaches $E_{c}$ from above, and from Eq.~(\ref{eq:CHI_MV})
we find $\chi^{(1)}_{zz} \propto (E_{c}-|E_{f}|)^{-1}$ as $|E_{f}|$
approaches $E_{c}$ from below. Thus, the critical exponents
according to mean-field theory are $\gamma=\gamma'=1$.

\section{Linear optical response}
\label{sec:CHIONE}

We first consider the linear optical response of the mixed-valent
system. The SCMF solution predicts an energy gap $2\Delta$ in
the absorption spectrum. The gap is $2\Delta$ because the
incoming photon must create {\em two} quasiparticles, just as
in a superconductor. Far-infrared transmission and reflectivity
spectra~\cite{Wachter,Molnar,Batlogg}, as well as electron
tunneling spectra~\cite{Guntherodt}, show energy gaps of
several meV in a number of mixed-valent compounds.
The crucial difference between the mixed-valent compound and the
superconductor is this: in the superconductor, the pairing occurs
between two {\em electrons}, whereas in the mixed-valent compound
the pairing occurs between an {\em electron} and a {\em hole}.
This has important consequences for the coherence factors that
enter the response of both systems to different external
probes. For example, it is a textbook result~\cite{Mahan} that a
clean superconductor at temperature $T=0$ cannot absorb
electromagnetic radiation because the coherence factor
$u_{\bf p}v_{\bf p+q}-v_{\bf p}u_{\bf p+q}$ vanishes for zero
photon momentum~{\bf q}. For the mixed-valent compound the coherence
factor entering the electromagnetic absorption is
$u_{\bf p}u_{\bf p+q}-v_{\bf p}v_{\bf p+q}$, which remains finite
for zero photon momentum. The coherence factor entering the
electromagnetic absorption of the mixed-valent compound is the
same as the coherence factor entering the {\em acoustic} attenuation
rate of the superconductor~\cite{Tinkham}.

We calculate the linear response of the mixed-valent system to
an ac electromagnetic field in the presence of a dc bias field.
The bias field serves to select the direction of the built-in
polarization. In a real material, the bias field is provided by
the crystal field or the depolarization field due to the sample
boundary. We treat the interaction of the mixed-valent system
with the ac electromagnetic field in the electric-dipole
approximation. The interaction term in the Hamiltonian is
$H_{\rm int}=-\mu_{z} {\cal E}_{z} \sum_{\bf k} d^{\dagger}_{\bf k}
f_{\bf k} + {\rm h.c.}$, where ${\cal E}_{z}$ is the component of
the ac electric field along the $z$\/-direction (i.e. the direction
of the built-in  polarization). Only the $z$\/-component of the ac
electric field couples to the channel in which the pairing takes
place. The optical signatures of $d$\/-$f$ exciton condensation
occur only in this channel. We ignore the response of the
remaining optical channels.

The ac electric field sets up a polarization $P_{z}$ in the material,
which in general can be a complicated nonlinear function of
${\cal E}_{z}$. In linear response, we expand $P_{z}$ in powers of
${\cal E}_{z}$ and keep only the first-order term: $P^{(1)}_{z} =
\chi^{(1)}_{zz} {\cal E}_{z}$. In the electric-dipole approximation,
the linear susceptibility $\chi^{(1)}_{zz}$ depends on the photon
frequency $\omega$ but not on the photon momentum ${\bf q}$. The
quantity measured in experiments is the reflectivity spectrum or
the transmission spectrum. From these one can extract the optical
conductivity $\sigma_{zz}(\omega)$ by Kramers-Kronig analysis. The
optical conductivity is related to the linear susceptibility by
$\sigma_{zz}(\omega)=-i\omega\chi^{(1)}_{zz}(\omega)$.

\subsection{Optical Bloch equations}

We have calculated the linear susceptibility $\chi^{(1)}_{zz}$
both from the Kubo formula and from the optical Bloch equations.
The pseudo-spin picture gives a nice physical description of the
linear and nonlinear responses of the system as  precessional modes
of the pseudo-spin vector ${\bf S}_{\bf k} =
(S_{x,{\bf k}},S_{y,{\bf k}},S_{z,{\bf k}})$. The optical Bloch
equations describe the time evolution of the pseudo-spin vector
under the action of the ac electric field ${\cal E}_{z}$. The
components of the pseudo-spin vector are the expectation values of
the pseudo-spin operators
\begin{eqnarray}
\sigma_{x,{\bf k}} & = & d^{\dagger}_{\bf k}f_{\bf k} +
                         f^{\dagger}_{\bf k}d_{\bf k} , \\
\sigma_{y,{\bf k}} & = & -i(d^{\dagger}_{\bf k}f_{\bf k} -
                            f^{\dagger}_{\bf k}d_{\bf k}) , \\
\sigma_{z,{\bf k}} & = & d^{\dagger}_{\bf k}d_{\bf k} -
                         f^{\dagger}_{\bf k}f_{\bf k}
\end{eqnarray}
in the ground state $|\psi\rangle$. The equations of motion for the
components of ${\bf S}_{\bf k}$ follow from the Heisenberg equations
of motion for the pseudo-spin operators ($\hbar=1$, $i=x,y,z$),
\begin{equation}
     \dot{\sigma}_{i,{\bf k}} = -i\,
    \left[ \sigma_{i,{\bf k}}, H + H_{\rm int} \right] .
\end{equation}
Working out the commutators, we find that the right-hand side
contains products of pseudo-spin operators
$\sigma_{i,{\bf k}} \sigma_{j,{\bf k'}}$ ($i\neq j$).
The products occur because of the Coulomb interaction term
in the Hamiltonian~(\ref{eq:FKH}). A closed set of
equations is obtained by replacing the average of products
$\langle\psi|\sigma_{i,{\bf k }}
             \sigma_{j,{\bf k'}}|\psi\rangle$
by the product of averages
$\langle\psi|\sigma_{i,{\bf k }}|\psi\rangle
 \langle\psi|\sigma_{j,{\bf k'}}|\psi\rangle =
S_{i,{\bf k}} S_{j,{\bf k'}}$. This gives the optical
Bloch equations
\begin{equation}
\label{eq:BLOCH}
\dot{\bf S}_{\bf k} =
( {\bf H}_{\bf k} - {\bf M}_{\bf k} ) \times {\bf S}_{\bf k} ,
\end{equation}
where ${\bf H}_{\bf k}=(-2\mu_{z}(E_{z}+{\cal E}_{z}),0,
\varepsilon_{\bf k}-E'_{f})$ and ${\bf M}_{\bf k}=\frac{U}{N}\sum_{\bf k}
{\bf S}_{\bf k}$. The symbol $\times$ represents the vector cross product.
The optical Bloch equations describe the coupled motion of a
collection of $N$ pseudo-spins. Each pseudo-spin precesses around
a local ``magnetic'' field ${\bf H}_{\bf k}-{\bf M}_{\bf k}$, which
is the sum of an external field ${\bf H}_{\bf k}$ and an average
internal field $-{\bf M}_{\bf k}$, where ${\bf M}_{\bf k}$ is the
pseudo-magnetization.

\subsection{Stationary solution}
\label{sec:STAT}
In the absence of the ac electric field the optical
Bloch equations have a stationary solution ${\bf S}^{(0)}_{\bf k}$.
The stationary solution is obtained by setting
$\dot{\bf S}^{(0)}_{\bf k} = 0$ in Eq.~(\ref{eq:BLOCH}):
\begin{equation}
\label{eq:STAT}
0 = ({\bf H}^{(0)}_{\bf k}-
     {\bf M}^{(0)}_{\bf k}) \times {\bf S}^{(0)}_{\bf k} .
\end{equation}
Here ${\bf H}^{(0)}_{\bf k} = (-2\mu_{z}E_{z},0,\varepsilon_{\bf k}-E'_{f})$,
and  ${\bf M}^{(0)}_{\bf k} = \frac{U}{N} \sum_{\bf k}
      {\bf S}^{(0)}_{\bf k}$.
In the stationary state each pseudo-spin is lined up with the
local ``magnetic'' field. Then there are two possibilities:
${\bf S}^{(0)}_{\bf k}$ is either parallel or antiparallel to
${\bf H}^{(0)}_{\bf k}-{\bf M}^{(0)}_{\bf k}$.
The state with ${\bf S}^{(0)}_{\bf k}$ antiparallel to
${\bf H}^{(0)}_{\bf k}-{\bf M}^{(0)}_{\bf k}$ has the lower energy.
Thus in the ground state all pseudo-spins point in the direction
opposite the local ``magnetic'' field.

For zero bias field ($E_{z}=0$), Eq.~(\ref{eq:STAT}) is invariant
under rotation about the $z$-axis. If ${\bf S}^{(0)}_{\bf k}$ is a
solution, then so is $R_{z}(\phi) {\bf S}^{(0)}_{\bf k}$, where
$R_{z}(\phi)$ is a rotation about the $z$-axis over an angle $\phi$.
The angle $\phi$ is the phase of the gap parameter $\Delta$. For
nonzero bias field, ${\bf S}^{(0)}_{\bf k}$ lies in the $x$-$z$ plane
and the gap parameter is real. Introducing spherical polar coordinates,
${\bf S}^{(0)}_{\bf k}=(\sin\theta_{\bf k},0,\cos\theta_{\bf k})$
and ${\bf M}^{(0)}_{\bf k}=(2\Delta,0,U n)$. The tilting angle
$\theta_{\bf k}$ is the angle between ${\bf S}^{(0)}_{\bf k}$ and
the positive $z$-axis. The magnetization ${\bf M}_{\bf k}^{(0)}$
must be determined self-consistently from ${\bf M}_{\bf k}^{(0)} =
\frac{U}{N}\sum_{\bf k}{\bf S}^{(0)}_{\bf k}$. The $z$-component of
this equation gives Eq.~(\ref{eq:FOCC}), and the $x$-component gives
Eq.~(\ref{eq:GAPEZ}). The stationary solution of the optical Bloch
equations is the SCMF solution given in section~\ref{sec:MODEL}.

\subsection{Pseudo-spin precession}
To calculate the linear susceptibility from the optical Bloch
equations we expand the pseudo-spin vector and
pseudo-magnetization to first order in ${\cal E}_{z}$:
${\bf S}_{\bf k}={\bf S}^{(0)}_{\bf k}+{\bf S}^{(1)}_{\bf k}$ and
${\bf M}_{\bf k}={\bf M}^{(0)}_{\bf k}+{\bf M}^{(1)}_{\bf k}$. Also,
${\bf H}_{\bf k}={\bf H}^{(0)}_{\bf k}+{\bf H}^{(1)}_{\bf k}$,
with ${\bf H}^{(1)}_{\bf k} = (-2\mu_{z}{\cal E}_{z},0,0)$.
We substitute the expansions into the Bloch equations and
collect terms of the same order in the ac electric field.
To zeroth order we recover Eq.~(\ref{eq:STAT}).
To first order we get
\begin{equation}
\label{eq:MODES}
\dot{\bf S}^{(1)}_{\bf k} -
   ({\bf H}^{(0)}_{\bf k} - {\bf M}^{(0)}_{\bf k}) \times
    {\bf S}^{(1)}_{\bf k} - {\bf S}^{(0)}_{\bf k}  \times
    {\bf M}^{(1)}_{\bf k} =
{\bf H}^{(1)}_{\bf k} \times {\bf S}^{(0)}_{\bf k} .
\end{equation}
Taking the inner product with the stationary solution gives
$\dot{\bf S}^{(1)}_{\bf k}\cdot{\bf S}^{(0)}_{\bf k}=0$.
The ac electric field causes the pseudo-spin vector to {\em precess}
about the stationary direction. With ${\bf S}^{(0)}_{\bf k}$
tilted away from the $z$-axis, the precession involves variations
in all three cartesian components of ${\bf S}_{\bf k}$.
The problem is simplified by working in the spherical polar
coordinate system. In spherical polar coordinates, the stationary
solution is the unit vector in the radial direction ${\bf e}_{r}$.
The precession about ${\bf e}_{r}$ is decomposed into components
along the polar and azimuthal units vectors ${\bf e}_{\theta}$
and ${\bf e}_{\phi}$:
\begin{math}
{\bf S}^{(1)}_{\bf k}=S^{(1)}_{\theta,\bf k} {\bf e}_{\theta} +
                      S^{(1)}_{\phi,  \bf k} {\bf e}_{\phi  } .
\end{math}
\mbox{ } One must remember that ${\bf e}_{r}$ and ${\bf e}_{\theta}$ vary
with the tilt angle $\theta_{\bf k}$, while ${\bf e}_{\phi}$ is fixed
since ${\bf S}^{(0)}_{\bf k}$ lies in the $x$-$z$ plane. The equations
of motion for $S^{(1)}_{\theta,\bf k}$ and $S^{(1)}_{\phi,\bf k}$ are
\begin{eqnarray}
\label{eq:MODE_1}
\dot{S}^{(1)}_{\theta,\bf k} - 2 E_{\bf k} S^{(1)}_{\phi  ,\bf k} +
      M^{(1)}_{\phi  ,\bf k} & = & 0 \\
\label{eq:MODE_2}
\dot{S}^{(1)}_{\phi  ,\bf k} + 2 E_{\bf k} S^{(1)}_{\theta,\bf k} -
      M^{(1)}_{\theta,\bf k} & = & 2 \mu_{z}{\cal E}_{z} \cos \theta_{\bf k} .
\end{eqnarray}
Here $M^{(1)}_{\phi  ,\bf k} =
     \frac{U}{N} \sum_{\bf k'} S^{(1)}_{\phi  ,\bf k'}$
and  $M^{(1)}_{\theta,\bf k} =
     \frac{U}{N} \sum_{\bf k'} \cos(\theta_{\bf k}-\theta_{\bf k'})
      S^{(1)}_{\theta,\bf k'}$.
The appearance of the cosine factor in $M^{(1)}_{\theta,\bf k}$
is due to the variation of ${\bf e}_{\theta}$ with $\theta_{\bf k}$.
In deriving Eqs.~(\ref{eq:MODE_1}) and (\ref{eq:MODE_2}) from
Eq.~(\ref{eq:MODES}) we have used the fact that $S^{(1)}_{r,\bf k} = 0$,
which follows from $\dot{S}^{(1)}_{r,\bf k} = 0$ and the initial condition
$S^{(1)}_{r,\bf k}(t=0) = 0$.

\subsection{Noninteracting quasiparticles}

It is instructive to first calculate the absorption spectrum
neglecting the Coulomb interaction between the optically
excited quasiparticles. In the pseudo-spin picture, this corresponds
to setting $M^{(1)}_{\phi,\bf k} = M^{(1)}_{\theta,\bf k} = 0$ in
Eqs.~(\ref{eq:MODE_1}) and (\ref{eq:MODE_2}). The linear susceptibility
is $\chi^{(1)}_{zz} = P^{(1)}_{z}/{\cal E}_{z}$, where
$P^{(1)}_{z} = (N \mu_{z}/\Omega) \sum_{\bf k} S^{(1)}_{\theta, \bf k}
\cos \theta_{\bf k}$ is the first-order polarization. Solving
Eqs.~(\ref{eq:MODE_1}) and (\ref{eq:MODE_2}) for $S^{(1)}_{\theta,\bf k}$,
we obtain
\begin{equation}
\label{eq:CHIONE}
\chi^{(1)}_{zz} = -\frac{N \mu^{2}_{z}}{\Omega} \sum_{\bf k}
\left(
\frac{n^{2}_{\bf k,k}}{\omega-2 E_{\bf k}} -
\frac{n^{2}_{\bf k,k}}{\omega+2 E_{\bf k}}
\right) ,
\end{equation}
where $n_{\bf k,k'}=u_{\bf k}u_{\bf k'}-v_{\bf k}v_{\bf k'}$ is a
coherence factor. The frequency $\omega$ is understood to have a small
positive imaginary part $\delta$. The physical origin of the coherence
factor $n_{\bf k,k'}$ can be understood as follows~\cite{Schrieffer}:
In the absorption process, an incoming photon
of momentum ${\bf k}-{\bf k'}$ creates a quasi-electron of momentum
${\bf k}$ and a quasi-hole of momentum $-{\bf k'}$. This can be done
in two different ways: (1) If the pair states $({\bf k},-{\bf k})$ and
$({\bf k'},-{\bf k'})$ are initially empty, by adding a $d$\/-electron
in ${\bf k}$ and adding an $f$\/-hole in $-{\bf k'}$. This process has
amplitude $u_{\bf k} u_{\bf k'}$. (2) If the pair states $({\bf k},-{\bf k})$
and $({\bf k'},-{\bf k'})$ are initially occupied, by removing an $f$\/-hole
from $-{\bf k}$ and removing a $d$\/-electron from ${\bf k'}$. This process
has amplitude $-v_{\bf k}v_{\bf k'}$. The overall amplitude for the
creation process is $u_{\bf k}u_{\bf k'}-v_{\bf k}v_{\bf k'}=n_{\bf k,k'}$.

For the model system the absorption spectrum can be found analytically.
When the $f$\/-level lies inside the $d$\/-band ($|E_{f}|\leq W$), the
energy gap is $2\Delta$, and the absorption rises as $\sqrt{\omega-2\Delta}$
above threshold. When the $f$\/-level lies outside the $d$\/-band
($|E_{f}|>W$), the gap is $\sqrt{(|E_{f}|-W)^{2}+4\Delta^{2}}$, and there
is a discontinuous jump in the absorption at threshold. The
single-quasiparticle result for the absorption spectrum at half-filling
($E_{f}=0$) is shown by the dash-dotted line in Fig.~\ref{fig:SMB}.

\subsection{Final-state interaction}

The quasiparticles created in the optical transition interact via the
Coulomb potential~$U$. In the normal state ($\Delta=0$) the final-state
interaction leads to the Wannier exciton. In the mixed-valent state
($\Delta>0$) the final-state interaction leads to ferroelectric
resonance and a threshold singularity in the infrared absorption spectrum.
In the pseudo-spin picture, the final-state interaction corresponds to
$M^{(1)}_{\phi,\bf k}$ and $M^{(1)}_{\theta,\bf k}$. For a separable
Coulomb potential, Eqs.~(\ref{eq:MODE_1}) and (\ref{eq:MODE_2}) can be
solved analytically. The pseudo-spin components are given by
\begin{eqnarray}
\label{eq:SOL_1}
S^{(1)}_{\theta,\bf k} & = &    \mu_{z}{\cal E}_{z}
\left( \frac{\Gamma({\bf k}, \omega)}{\omega-2 E_{\bf k}} -
       \frac{\Gamma({\bf k},-\omega)}{\omega+2 E_{\bf k}} \right) , \\
\label{eq:SOL_2}
S^{(1)}_{\phi  ,\bf k} & = & -i \mu_{z}{\cal E}_{z}
\left( \frac{\Gamma({\bf k},\omega)}{\omega-2 E_{\bf k}} +
       \frac{\Gamma({\bf k},-\omega)}{\omega+2 E_{\bf k}} \right) ,
\end{eqnarray}
where
\begin{eqnarray}
\label{eq:VERTEX}
\Gamma({\bf k},\omega) & = & -\cos \theta_{\bf k}
-\frac{U}{N} \sum_{\bf k'} \frac{
\cos^{2}( \frac{\theta_{k}-\theta_{k'}}{2} )
 \Gamma({\bf k'}, \omega)}{\omega-2 E_{\bf k'}}
\nonumber \\ & &
-\frac{U}{N} \sum_{\bf k'} \frac{
\sin^{2}( \frac{\theta_{k}-\theta_{k'}}{2} )
 \Gamma({\bf k'},-\omega)}{\omega+2 E_{\bf k'}}
\end{eqnarray}
is a vertex function. In diagrammatic terms, the vertex function is
the sum of all ladder diagrams contributing to the propagation of
the quasiparticle pair.

In the normal state ($\theta_{\bf k}=\pi$) Eq.~(\ref{eq:VERTEX}) yields
the Wannier exciton. The pole of the vertex function gives the exciton
binding energy $E_{\rm b}=W\coth(W/U)-W$. The absorption spectrum of the
normal state consists of an exciton line at $\omega=E_{g}-E_{b}$, and a
continuum between $\omega=E_{g}$ and $\omega=E_{g}+2W$, where
$E_{g}=-W-E_{f}$ is the $d$\/-$f$ band gap. The final-state interaction
enhances the absorption at $\omega=E_{g}$.

In the mixed-valent state the vertex function has three components:
\begin{equation}
\label{eq:COMPONENTS}
\Gamma({\bf k},\omega) = -\cos \theta_{\bf k} \Gamma_{1}(\omega)
 + \sin \theta_{\bf k} \Gamma_{2}(\omega) + \Gamma_{3}(\omega) .
\end{equation}
The components $\Gamma_{1}(\omega)$ and $\Gamma_{2}(\omega)$ are even
in $\omega$, while $\Gamma_{3}(\omega)$ is odd. Substitution of
Eq.~(\ref{eq:COMPONENTS}) into Eq.~(\ref{eq:VERTEX}) yields a set of
of three algebraic equations for the vertex components:
\begin{equation}
\label{eq:SET}
\left(
\begin{array}{ccc}
(\omega^{2}-M^{2})A(\omega)+R & M B(\omega) & \omega B(\omega) \\
M B(\omega) & 1 + M^{2} A(\omega) & \omega M A(\omega) \\
\omega B(\omega) & \omega M A(\omega) & \omega^{2} A(\omega)+R
\end{array}
\right) \left(
\begin{array}{c}
\Gamma_{1}(\omega) \\ \Gamma_{2}(\omega) \\ \Gamma_{3}(\omega)
\end{array}
\right) = \left(
\begin{array}{c}
1 \\ 0 \\ 0
\end{array} \right).
\end{equation}
Here $M=2\Delta+\mu_{z}E_{z}$ and $R=\mu_{z}E_{z}/(\Delta+\mu_{z}E_{z})$.
The functions $A(\omega)$ and $B(\omega)$ are given by
\begin{eqnarray}
\label{eq:FUNCA}
A(\omega) & = & \frac{U}{N} \sum_{\bf k}
\frac{1}{2 E_{\bf k}(\omega-2 E_{\bf k})
                    (\omega+2 E_{\bf k})} , \\
\label{eq:FUNCB}
B(\omega) & = & \frac{U}{N} \sum_{\bf k}
\frac{\varepsilon_{\bf k}-E_{f}}
       {2 E_{\bf k}(\omega-2 E_{\bf k})
                    (\omega+2 E_{\bf k})} .
\end{eqnarray}
For the model system $A(\omega)$ and $B(\omega)$ can found analytically.
The explicit expressions are given in the Appendix.
Solving Eq.~(\ref{eq:SET}) for $\Gamma_{1}(\omega)$ gives
\begin{equation}
\label{eq:GAMMA_1}
\Gamma_{1}(\omega) =
\frac{(\omega^{2}+R M^{2})A(\omega)+R}
    {[(\omega^{2}-  M^{2})A(\omega)+R]
     [(\omega^{2}+R M^{2})A(\omega)+R] -
      (\omega^{2}+R M^{2})B^{2}(\omega)} .
\end{equation}
The linear susceptibility is
\begin{equation}
\chi^{(1)}_{zz}(\omega) =
\frac{2N \mu_{z}^{2}}{\Omega U}(\Gamma_{1}(\omega) - 1).
\end{equation}
\subsection{Infrared absorption spectrum}

We first consider the absorption spectrum in the absence of a
static electric field.  We calculate the continuum absorption in
zero bias field, setting $M=2\Delta$ and $R=0$ in Eq.~(\ref{eq:GAMMA_1}).
For an energy gap of several meV, the continuum lies in the far
infrared.  The linear susceptibility in zero bias field is
\begin{equation}
\label{eq:CHIONE_2}
\chi^{(1)}_{zz}(\omega) = \frac{2 N \mu^{2}_{z}}{\Omega U}
\left( \frac{A(\omega)} {(\omega^{2}-4\Delta^{2})
       A^{2}(\omega)-B^{2}(\omega)}-1 \right).
\end{equation}
The solid lines in Fig.~\ref{fig:CHIONE} show the imaginary part of
Eq.~(\ref{eq:CHIONE_2}) for several values of $E_{f}$. When the $f$\/-level
lies far below the bottom of the $d$\/-band (top curve), the continuum
absorption of the mixed-valent state is very similar to the continuum
absorption of the normal state. There is a slight enhancement near the
energy gap.  As the $f$\/-level approaches the bottom of the
$d$\/-band, the enhancement becomes more and more pronounced. When the
$f$\/-level lies within the
$d$\/-band, the spectrum has a threshold singularity at $\omega=2\Delta$.
The spectrum for $E_{f}$ above the center of the $d$\/-band is the same
as the spectrum for $-E_{f}$ below it.
{}From Eq.~(\ref{eq:CHIONE_2}) we find that when $|E_{f}|<W$, the threshold
singularity is $\epsilon^{-1/2} \theta(\epsilon)$, and when $|E_{f}|=W$ the
singularity is $\epsilon^{-1/2} \ln^{-2}(\epsilon) \theta(\epsilon)$, where
$\epsilon=\omega-2\Delta$. When the $f$\/-level lies outside the $d$\/-band
the singularity is cut off because the energy gap is larger than $2\Delta$.
The singularity is due to the final-state interaction between the optically
excited quasiparticles. In the single-quasiparticle treatment, the absorption
rises continuously from zero according to $\epsilon^{1/2}\theta(\epsilon)$.
The singularity is {\em not} an artifact of the simple model, and should be
observable in real materials. Roundoff may occur due to lifetime effects
and sample inhomogeneities.

\subsection{Ferroelectric resonance}

The dash-dotted line in Fig.~\ref{fig:CHIONE} represents the
absorption spectrum of the mixed-valent system in a bias field.
The spectrum consists of two parts: a peak at $\omega=\omega_{0}$,
and a continuum above $\omega=2\Delta$. We shall show that
 $\omega_{0}$ is the ferroelectric resonance frequency.
 The continuum part of the absorption spectrum
is blue-shifted only slightly by the bias field.

Ferroelectric resonance occurs when an electronic ferroelectric,
placed in a bias field $E_{z}$, is acted upon by an alternating
field ${\cal E}_{z}$ of frequency $\omega_{0}$. The ferroelectric
resonance frequency $\omega_{0}$ is the frequency of the Goldstone
mode in the bias field $E_{z}$. The Goldstone mode corresponds to a
uniform precession of the pseudo-spins around the $z$\/-axis. For an
ideal system $\omega_{0}$ vanishes at zero bias field. For a real
system the external bias field $E_{z}$ must be replaced by an effective
internal field $E_{\rm eff}$, which may depend on the sample shape, the
domain structure, and the crystal anisotropy. An example of a contribution
to $E_{\rm eff}$ is the depolarization field $-N_{z}P_{z}$ of the sample
boundary, where $N_{z}$ is the depolarization factor. The effective field
remains nonzero in the absence of the external bias field, yielding a
finite resonance frequency for an unbiased sample.

Ferroelectric resonance is the electric analogue of
ferromagnetic resonance in a ferromagnetic insulator. In the
magnetic case, the alternating field causes a uniform precession of
{\em real} spins around the $z$\/-direction. The ferromagnetic resonance
frequency is $\omega_{0}=\gamma H_{z}$, where $\gamma$ is
the gyromagnetic ratio, and $H_{z}$ is the bias field. In real samples
$H_{z}$ must be replaced by an effective field $H_{\rm eff}$ depending
on the sample geometry, magnetic domain structure, and crystal magnetic
anisotropy~\cite{Vonsovskii}.

Ferroelectric resonance does not occur in displacive ferroelectrics
because the order parameter only has a discrete symmetry. This
means there is no Goldstone mode in the excitation spectrum of a
displacive ferroelectric. Ferroelectric liquid crystals {\em do}
have an order parameter with a continuous symmetry. The dielectric
response of the Goldstone mode has been observed~\cite{Zubia}
in ferroelectric liquid crystals by means of broadband dielectric
spectroscopy.

We now calculate the dependence of the ferroelectric resonance frequency
of an ideal system on the external bias field $E_{z}$. For a real system,
$E_{z}$ must be replaced by $E_{\rm eff}$. The ferroelectric resonance
frequency is given by the equation $D(\omega_{0})=0$,
where $D(\omega)$ is the denominator of Eq.~(\ref{eq:GAMMA_1}).
For $\mu_{z}E_{z}$ small compared to $\Delta$, an approximate
solution may be obtained by expanding $D(\omega_{0})$ in a Taylor
series around $\omega_{0}=0$:
     $D(\omega_{0})=D(0)+\frac{1}{2}\omega_{0}^{2}D''(0)$.
The linear term vanishes because $D(\omega)$ is even in $\omega$.
Neglecting terms of order $R^{2}$ in $D(0)$ and of order $R$ in $D''(0)$,
we find
\begin{equation}
\omega_{0} = \left[ -4 \mu_{z} \Delta \left( 1 + \frac{A(0)}
{4\Delta^{2}A^{2}(0) + B^{2}(0)} \right) \right]^{1/2} \sqrt{E_{z}}.
\end{equation}
Substituting the explicit expressions for $A(0)$ and $B(0)$ given
in the Appendix yields
\begin{equation}
\label{eq:WRES}
\omega_{0} = \left( \frac{4 \mu_{z}\Delta E'_{c}}{U} \right)^{1/2}
\sqrt{E_{z}},
\end{equation}
where $E'_{c}=E_{c}-U$ is the critical value of $E'_{f}$. A useful
estimate of $\omega_{0}$ is the arithmetic mean of the gap $2\Delta$
and the field energy $\mu_{z}E_{z}$. For $\mu_{z}=10^{-29}$ Cm,
$2\Delta=1$ meV, and $E_{z}$ between $10$ V/cm and $10^{4}$ V/cm,
$\omega_{0}$ is between $0.01$ meV and $1$ meV, i.e. in the microwave
regime.

The condition $D(\omega_{0})=0$ is not sufficient for a peak in
the absorption spectrum at $\omega=\omega_{0}$. One must also have
$N(\omega_{0}) \neq 0$, where $N(\omega)$ is the numerator of
Eq.~(\ref{eq:GAMMA_1}). The strength of the pole at $\omega_{0}$ is
$Z_{0}=-\pi N(\omega_{0})/D'(\omega_{0})$. Away from half-filling,
one finds $Z_{0}>0$. Exactly at half-filling, $B(\omega)=0$. As can
be seen at once from Eq.~(\ref{eq:GAMMA_1}), we then have $Z_{0}=0$.
The zero strength at half-filling is an artifact of our simple model,
which has $\rho(\epsilon)=\rho(-\epsilon)$. For a realistic $d$\/-band
there will be a resonance peak at half-filling.

An important problem in ferromagnetic resonance is to account for the
width of the ferromagnetic resonance line~\cite{Sparks}. The line width
is due to the relaxation of the uniform precession by spin-spin and
spin-lattice interactions. Various damping terms may be added to the
Bloch equations to describe the relaxation. A possible damping term
is $-({\bf S}_{\bf k}-{\bf S}^{(0)}_{\bf k})/\tau$, where $\tau$ is a
phenomenological relaxation time. In the equation of motion for
$S^{(1)}_{\bf k}$, this leads to the replacement of $d/dt$
by $d/dt+1/\tau$.
Thus, the simplest approach to damping (the one we have adopted here) is
 to add an imaginary part $\delta = 1/\tau$ to $\omega$.
A detailed study of the width of the ferroelectric resonance line is
left for future research.

\section{Second-harmonic generation}
\label{sec:CHITWO}

Second-harmonic generation is the generation of an outgoing
electromagnetic wave of frequency $2\omega$ from two incoming
waves of frequency $\omega$. For incoming waves propagating
along ${\bf k}_{1}$ and ${\bf k}_{2}$, the second-harmonic
radiation is most effectively generated in the phase-matching
direction ${\bf k}_{1}+{\bf k}_{2}$. In this direction both
energy and momentum are conserved.

The ability of a medium to sustain second-harmonic generation
is characterized by the second-harmonic susceptibility tensor
$\chi^{(2)}_{ijk}(2\omega,\omega,\omega)$. Here $i,j,k=x,y,z$
are the Cartesian indices. The polarization $P_{i}(2\omega)$
induced in the medium by incoming fields ${\cal E}_{j}(\omega)$ and
${\cal E}_{k}(\omega)$ is given by $P_{i}(2\omega)=\chi^{(2)}_{ijk}
(2\omega,\omega,\omega){\cal E}_{j}(\omega){\cal E}_{k}(\omega)$.
For a medium with inversion symmetry, the second-harmonic
susceptibility must satisfy $\chi^{(2)}_{ijk}=-\chi^{(2)}_{ijk}=0$.
Therefore, second-harmonic generation cannot occur in media with
inversion symmetry.

In the mixed-valent system, the inversion symmetry
 is spontaneously broken by the pairing of electronic states
of opposite parity. This leads to the appearance of a built-in
polarization in the $z$\/-direction. If the incoming fields have
a component along the $z$\/-axis, second-harmonic generation can occur.
We calculate the second-harmonic susceptibility $\chi^{(2)}_{zzz}$ for
incoming fields polarized along the $z$\/-axis. For incoming fields
polarized at an angle $\phi$ relative to the $z$\/-axis, the
second-harmonic susceptibility is reduced by $\cos^{2}\phi$.

\subsection{Pseudo-spin nutation}

We calculate the second-harmonic susceptibility $\chi^{(2)}_{zzz}$
from the optical Bloch equations by expanding ${\bf S}_{\bf k}$ and
${\bf M}_{\bf k}$ to second order in ${\cal E}_{z}$. The external
``magnetic'' field ${\bf H}_{\bf k}$ has no components of second or
higher order. The equation of motion for ${\bf S}^{(2)}_{\bf k}$ is
\begin{equation}
\label{eq:SAMES}
\dot{\bf S}^{(2)}_{\bf k} -
  ( {\bf H}^{(0)}_{\bf k} - {\bf M}^{(0)}_{\bf k} ) \times
    {\bf S}^{(2)}_{\bf k} + {\bf M}^{(2)}_{\bf k}   \times
    {\bf S}^{(0)}_{\bf k} =
  ( {\bf H}^{(1)}_{\bf k} - {\bf M}^{(1)}_{\bf k} ) \times
    {\bf S}^{(1)}_{\bf k} .
\end{equation}
Eq.~(\ref{eq:SAMES}) has the same form as Eq.~(\ref{eq:MODES}),
except with a more complicated right hand side. As before, we
decompose ${\bf S}^{(2)}_{\bf k}$ into its radial, polar, and
azimuthal components: ${\bf S}^{(2)}_{\bf k} = S^{(2)}_{r,\bf k}
{\bf e}_{r} + S^{(2)}_{\theta,\bf k} {\bf e}_{\theta} +
S^{(2)}_{\phi,\bf k} {\bf e}_{\phi}$. To second order, the pseudo-%
spin has a nonzero radial component $S^{(2)}_{r,\bf k}$. This means
the motion is no longer a regular precession: the pseudo-spin
{\em nutates} during the precession. (Nutation is the up-and-down motion
of the precession axis.) The nutation frequency is twice the precession
frequency. This can be shown as follows: the equation of motion for
$S^{(2)}_{r,\bf k}$ is obtained by taking the inner product of
Eq.~(\ref{eq:SAMES}) with ${\bf e}_{r}$. Using the vector identity
$({\bf A} \times {\bf B}) \cdot {\bf C} = ({\bf C} \times {\bf A}) \cdot
{\bf B}$ and taking the inner product of Eq.~(\ref{eq:MODES}) with
${\bf e}_{r}$ shows that the equation for $S^{(2)}_{r,\bf k}$ can be
written as $\dot{S}^{(2)}_{r,\bf k} = -\dot{\bf S}^{(1)}_{\bf k} \cdot
{\bf S}^{(1)}_{\bf k}$. Upon integration, we find
\begin{equation}
\label{eq:NUTATION}
S^{(2)}_{r,\bf k} = -\frac{1}{2} (S^{(1)}_{\theta,\bf k}
S^{(1)}_{\theta,\bf k} + S^{(1)}_{\phi,\bf k} S^{(1)}_{\phi,\bf k}) .
\end{equation}

The equations of motion for $S^{(2)}_{\theta,\bf k}$ and
$S^{(2)}_{\phi,\bf k}$ are obtained by taking the inner product
of Eq.~(\ref{eq:SAMES}) with ${\bf e}_{\theta}$ and ${\bf e}_{\phi}$
respectively. Apart from the more complicated source terms, a new
feature occuring in second order is that $S^{(2)}_{r,\bf k}$ is now
nonzero. This leads to an additional term $\frac{U}{N} \sum_{\bf k'}
\sin(\theta_{\bf k}-\theta_{\bf k'}) S^{(2)}_{r,\bf k'}$ on the left
hand side of the equation for $S^{(2)}_{\phi,\bf k}$. However, since we
have already solved for $S^{(2)}_{r,\bf k}$ in Eq.~(\ref{eq:NUTATION}),
the additional term can be taken over to the right hand side and treated
as an extra source term. Combining all source terms into driving forces
$F^{(2)}_{\theta,\bf k}$ and $F^{(2)}_{\phi,\bf k}$, the equations of
motion for $S^{(2)}_{\theta,\bf k}$ and $S^{(2)}_{\phi,\bf k}$ are
\begin{eqnarray}
\label{eq:SAME_1}
\dot{S}^{(2)}_{\theta,\bf k} - 2 E_{\bf k} S^{(2)}_{\phi  ,\bf k} +
      M^{(2)}_{\phi  ,\bf k} & = & F^{(2)}_{\theta,\bf k} \\
\label{eq:SAME_2}
\dot{S}^{(2)}_{\phi  ,\bf k} + 2 E_{\bf k} S^{(2)}_{\theta,\bf k} -
      M^{(2)}_{\theta,\bf k} & = & F^{(2)}_{\phi,\bf k} ,
\end{eqnarray}
with
\begin{eqnarray}
\label{eq:FORCE_1}
F^{(2)}_{\theta,\bf k} & = &
 (2 \mu_{z}{\cal E}_{z} \sin\theta_{\bf k} + M^{(1)}_{r,\bf k})
S^{(1)}_{\phi,\bf k} , \\
\label{eq:FORCE_2}
F^{(2)}_{\phi,\bf k} & = &
-(2 \mu_{z}{\cal E}_{z} \sin\theta_{\bf k} + M^{(1)}_{r,\bf k})
S^{(1)}_{\theta,\bf k} - \frac{U}{N} \sum_{\bf k'}
\sin(\theta_{\bf k}-\theta_{\bf k'}) S^{(2)}_{r,\bf k'} .
\end{eqnarray}
Here $M^{(1)}_{r,\bf k} = \frac{U}{N} \sum_{\bf k'}
\sin(\theta_{\bf k}-\theta_{\bf k'}) S^{(1)}_{\theta,\bf k'}$.
A very important
observation is that since $\sin\theta_{\bf k} = \Delta/E_{\bf k}$,
all source terms are proportional to $\Delta$.
The second-harmonic susceptibility vanishes identically when
$\Delta=0$.

\subsection{Independent quasiparticles}

We first calculate the second-harmonic
susceptibility neglecting the Coulomb interaction between the
optically excited quasiparticles. This corresponds to setting
$M^{(2)}_{\theta,\bf k} = M^{(2)}_{\phi,\bf k} = 0$ on the left
hand side of Eqs.~(\ref{eq:SAME_1}) and (\ref{eq:SAME_2}), and
$F^{(2)}_{\theta,\bf k} = 2\mu_{z}{\cal E}_{z} S^{(1)}_{\phi,\bf k}
\sin \theta_{\bf k}$ and $F^{(2)}_{\phi,\bf k} =-2\mu_{z}{\cal E}_{z}
S^{(1)}_{\theta,\bf k} \sin\theta_{\bf k}$ on the right hand side.
The second-harmonic susceptibility is $\chi^{(2)}_{zzz} =
P^{(2)}_{z}/{\cal E}^{2}_{z}$, where $P^{(2)}_{z} = (N \mu_{z}/\Omega)
\sum_{\bf k}(S^{(2)}_{\theta,\bf k} \cos\theta_{\bf k} +
S^{(2)}_{r,\bf k} \sin \theta_{\bf k})$ is the second-order
polarization. The second-order polarization has a
contribution from the radial component $S^{(2)}_{r,\bf k}$.
Solving Eqs.~(\ref{eq:SAME_1}) and (\ref{eq:SAME_2})
for $S^{(2)}_{\theta,\bf k}$, and using Eq.~(\ref{eq:NUTATION}) for
$S^{(2)}_{r,\bf k}$, we obtain
\begin{eqnarray}
\label{eq:CHITWO}
\chi^{(2)}_{zzz} & = &
- \frac{N \mu_{z}^{3}}{\Omega} \sum_{\bf k}
\left(
\frac{2 m_{\bf k,k} n_{\bf k,k}^{2} }
     {(\omega - 2E_{\bf k})(2\omega - 2 E_{\bf k})}
\right. \nonumber \\ & & \left.
+ \frac{2 m_{\bf k,k} n_{\bf k,k}^{2} }
     {(\omega + 2E_{\bf k})(2\omega + 2 E_{\bf k})}
- \frac{2 m_{\bf k,k} n_{\bf k,k}^{2} }
     {(\omega - 2E_{\bf k})(\omega + 2 E_{\bf k})}
\right) .
\end{eqnarray}
The coherence factor $m_{\bf k,k'}$ is given by $m_{\bf k,k'} =
u_{\bf k}v_{\bf k'} + v_{\bf k}u_{\bf k'}$. The physical origin
of the coherence factor $m_{\bf k,k'}$ is the scattering of a
quasiparticle by the second incoming photon. The second photon
can either scatter the quasi-electron from ${\bf k}$ to ${\bf k'}$,
or the quasi-hole from $-{\bf k'}$ to $-{\bf k}$. The overall
amplitude for the scattering process is $2m_{\bf k,k'}$. Since
$m_{\bf k,k}$ = $\Delta/E_{\bf k}$, the second-harmonic susceptibility
is directly proportional to $\Delta$.

\subsection{Final-state interaction}

For a separable Coulomb potential, an analytic solution for the
second-harmonic susceptibility including the final-state interaction
is possible in principle. However, the large number of driving terms
in Eqs.~(\ref{eq:FORCE_1}) and (\ref{eq:FORCE_2}) presents a
considerable challenge. We have instead approached the problem
numerically. This is done in analogy with the classical mechanics
treatment of forced oscillations. The azimuthal component
$S_{\phi,\bf k}$ is the generalized coordinate $q_{\bf k}$, and
the polar component $S_{\theta,\bf k}$ is the negative of the conjugate
momentum $p_{\bf k}$. Eqs.~(\ref{eq:SAME_1}) and (\ref{eq:SAME_2})
are the Hamilton equations of motion for $q_{\bf k}$ and $p_{\bf k}$.
The matrices $T$ and $V$ are
\begin{eqnarray}
T_{\bf k,k'}^{-1} & = &
2 E_{\bf k} \delta_{\bf k,k'} - \frac{U}{N}
\cos(\theta_{\bf k}-\theta_{\bf k'}) , \\
V_{\bf k,k'}      & = &
2 E_{\bf k} \delta_{\bf k,k'} - \frac{U}{N} .
\end{eqnarray}
The first step is to find the normal modes of oscillation of
the system of pseudo-spins. The normal-mode equations are
\begin{equation}
\sum_{\bf k'} V_{\bf k,k'} A_{{\bf k'},n} = (2 E_{n})^{2}
\sum_{\bf k'} T_{\bf k,k'} A_{{\bf k'},n} ,
\end{equation}
where $2E_{n}$ is the frequency of normal mode~$n$, and
$A_{{\bf k},n}$ is the amplitude of $S_{\phi,\bf k}$ in the
normal mode~$n$. Since $T$ and $V$ are both real and symmetric, the
frequencies are all real and positive. There is one Goldstone
mode, whose frequency is the ferroelectric resonance
frequency~$\omega_{0}$.
The remaining $N-1$ frequencies form a continuum above the energy gap.

The second step is to obtain the forced oscillation of
$S^{(2)}_{\theta,\bf k}$ and $S^{(2)}_{\phi,\bf k}$ when driven by
$F^{(2)}_{\theta,\bf k}$ and $F^{(2)}_{\phi,\bf k}$.
This is done by solving for the motion in normal coordinates, and
then taking linear combinations to obtain the motion in the original
coordinates. The force driving the normal coordinate $\zeta_{n}$ has
frequency $2\omega$ and amplitude $Q_{n}=-\sum_{\bf k} (A_{{\bf k},n}
F^{(2)}_{\theta,\bf k}+2i\omega A_{n,\bf k}^{-1}F^{(2)}_{\phi,\bf k})$.
This causes the normal coordinate to oscillate with frequency
$2\omega$ and amplitude $\zeta_{n} = Q_{n}/(4E_{n}^{2}-4\omega^{2})$.
The original coordinates oscillate with frequency $2\omega$ and
amplitudes $S^{(2)}_{\phi,\bf k} = \sum_{n} A_{{\bf k},n}\zeta_{n}$
and $S^{(2)}_{\theta,\bf k} = 2 i \omega \sum_{n} A_{{\bf k},n}^{-1}
\zeta_{n} + \sum_{\bf k'} T_{\bf k,k'} F^{(2)}_{\phi,\bf k'}$. The
complete motion in the original coordinates is ($m=2$):
\begin{eqnarray}
\label{eq:PIECES_1}
S^{(m)}_{\phi,\bf k} & = & \sum_{n,\bf k'} \frac{
A_{{\bf k},n}(A_{{\bf k'},n}  F^{(m)}_{\theta,\bf k'}
+m i \omega A_{n,\bf k'}^{-1} F^{(m)}_{\phi,\bf k'} )}
{m^{2} \omega^{2}-4 E_{n}^{2}} , \\
\label{eq:PIECES_2}
S^{(m)}_{\theta,\bf k} & = &  \sum_{n,\bf k'} \frac{
A_{n,\bf k}^{-1}( 4 E_{n}^{2} A_{n,\bf k'}^{-1} F^{(m)}_{\phi,\bf k'}
-m i \omega A_{{\bf k'},n} F^{(m)}_{\theta,\bf k'} )}
{4 E_{n}^{2}- m^{2} \omega^{2}} .
\end{eqnarray}
Here we have used the orthonormality condition $\delta_{n,n'} =
\sum_{\bf k,k'} A_{{\bf k},n}T_{\bf k,k'}A_{{\bf k'},n'}$ for the
matrix of eigenvectors $A_{{\bf k},n}$. Eqs.~(\ref{eq:PIECES_1}) and
(\ref{eq:PIECES_2}) apply to $m$\/-th harmonic generation in general.

The computational task is summarized as follows: First compute
$S^{(1)}_{\phi,\bf k}$ and $S^{(1)}_{\theta,\bf k}$ from
Eqs.~(\ref{eq:PIECES_1}) and (\ref{eq:PIECES_2}) with $m=1$
and $F^{(1)}_{\theta,\bf k} = 0$, $F^{(1)}_{\phi,\bf k} =
2\mu_{z}{\cal E}_{z} \cos\theta_{\bf k}$. To avoid a singular
denominator, $\omega$ is given a small positive imaginary part $\delta$.
Then compute $S^{(2)}_{r,\bf k}$ from Eq.~(\ref{eq:NUTATION}),
and $F^{(2)}_{\theta,\bf k}$ and $F^{(2)}_{\phi,\bf k}$ from
Eqs.~(\ref{eq:FORCE_1}) and (\ref{eq:FORCE_2}). Finally, compute
$S^{(2)}_{\theta,\bf k}$ from Eq.~(\ref{eq:PIECES_2}) with $m=2$.

The results of the calculation are shown in Fig.~\ref{fig:CHITWO}.
The figure shows the amplitude $|\chi^{(2)}_{zzz}(2\omega,\omega,\omega)|$
of the second-harmonic susceptibility as a function of the photon energy
$\omega$, for several values of $E_{f}$, in zero bias field. The important
features are: (1) The second-harmonic susceptibility is directly proportional
to $\Delta$. (2) When the $f$\/-level lies inside the $d$\/-band, the
second-harmonic conversion efficiency is strongly enhanced at $\omega=\Delta$,
and less strongly at $\omega=2\Delta$. The first feature shows that
second-harmonic generation can be used to test the validity of the SCMF
solution
in real mixed-valent compounds. The second feature distinguishes the
single-quasiparticle treatment of the second-harmonic response from
the SCMF treatment. The enhancement of the conversion efficiency is due to
the final-state interaction between the optically excited quasiparticles.

\section{Existing experimental evidence and proposed tests}
\label{sec:EXPERIMENT}

As experimental tests of
electronic ferroelectricity in a mixed-valent compound
we propose measurements of the static dielectric constant, the microwave
absorption spectrum, and the second-harmonic susceptibility.
As an example, consider SmB$_{6}$. The crystal structure of
SmB$_{6}$ has cubic symmetry, with B$_{6}$ octahedra at the body
center, and Sm ions at the corners of a conventional bcc unit
cell with lattice constant $a = 4.13\AA$. The crystal has
inversion symmetry at the bcc lattice points. Through measurements
of the ionic radius, the valence of the Sm ion is found to be 2.53,
almost halfway between 2 and 3, so that the $f$\/-level lies near
the center of the conduction band.

The measured far-infrared absorption spectrum~\cite{Wachter,Molnar,Batlogg}
of SmB$_{6}$ can be interpreted in accordance with the SCMF solution.
In Fig.~\ref{fig:SMB} we compare the mean-field and
single-quasiparticle results for the linear susceptibility to
experimental data on SmB$_{6}$ taken from Ref.~\onlinecite{Wachter}.
The data show an energy gap around $2\Delta=4$ meV, and a sharp peak
at threshold. The mean-field theory fits the data very well in the
threshold region, whereas the single-quasiparticle theory gives a
qualitatively wrong behaviour. Away from threshold, discrepancies
between mean-field theory and experiment occur because of our simple
model. Further experimental indication of the validity of the
SCMF solution in SmB$_{6}$ is provided by the electron tunneling
spectrum~\cite{Guntherodt}, which can be interpreted by analogy with
Giaever tunneling in a superconductor.

Ref.~\onlinecite{Wachter} also reports a measurement of the static
dielectric constant of SmB$_{6}$ at $T=4 K$. The large observed value
 $\epsilon=1,500$ provides further support for the theoretical
prediction of electronic ferroelectricity. An interesting test would
be the existence of the ferroelectric resonance in this compound.
 For the parameter values of Fig.~\ref{fig:SMB},
the ferroelectric resonance frequency is
$\omega_{0}=0.021\sqrt{E_{z}}$ meV, where $E_{z}$ is in V/cm.
This estimate from Eq.~(\ref{eq:WRES}) was checked by the numerical
solution of $D(\omega_{0})=0$ and was found to be in agreement to
 better than $0.5$\% for fields up to $10^{3}$ V/cm.
For a reasonably strong applied electric field, the ferroelectric
resonance lies easily in the range of frequency measured in
Ref.~\onlinecite{Wachter} at zero static electric field since the
lowest frequency measured there was $1$~meV.

Ref.~\onlinecite{Goto} reports measurements of the dielectric response
of mixed-valent Sm$_{3}$Se$_{4}$ and Sm$_{2}$Se$_{3}$ from $20$ Hz up
to $1$ GHz. The huge observed values of the static dielectric constant
($\epsilon=30,000$ for Sm$_{3}$Se$_{4}$ and $\epsilon=4,000$ for
Sm$_{2}$Se$_{3}$) are consistent with electronic ferroelectricity.
Given the value $2\Delta=140$ meV for the energy gap in Sm$_{3}$Se$_{4}$,
we predict a resonance frequency between $0.1$ meV ($10$ GHz) and
$10$ meV ($1$ THz).

Ref.~\onlinecite{Bucher} reports transport measurements on mixed-valent
TmSe$_{0.45}$Te$_{0.55}$ that show evidence for a condensation of free
carriers into an excitonic insulator ground state. Due to the indirect
nature of the energy gap, this material might not be suitable for the
optical tests of coherence proposed here.

\section{Discussion of other ground states}
\label{sec:DISCUSSION}

We now discuss the possibility of electronic ferroelectricity,
ferroelectric resonance, and second-harmonic generation for
solutions of the FK model {\em other} than the SCMF solution.
There are two types of solutions we consider here: (1) solutions
with a classical $f$\/-electron distribution, and (2) electronic-%
polaron solutions. The question we ask is: does the theoretical
ground state have a built-in coherence between $d$\/-electrons
and $f$\/-holes? This coherence is necessary for electronic
ferroelectricity and the concomitant optical signatures to occur.
The answer is, for the two types of solutions: (1) no, and (2) yes.
A brief motivation for each answer is given below.

In Refs.~\onlinecite{Freericks,Farkasovsky} the $f$\/-occupation
$f^{\dagger}_{i}f_{i}$ is replaced by a classical variable $W_{i}$,
where $W_{i}=1$ when site~$i$ is occupied by an $f$\/-electron and
$W_{i}=0$ when it is not. With this replacement, the FK model becomes
a tight-binding model with an on-site potential that can assume two
different values $U$ or $0$. For a given configuration of $f$\/-electrons
the ground state energy is found by filling the lowest $d$\/-electron
levels. The ground configuration is the configuration with the lowest
ground state energy. The question we ask is: what is the value of $\Delta=
\frac{U}{N}\sum_{i=1}^{N}\langle\psi|f^{\dagger}_{i}d_{i}|\psi\rangle$?
Since each site is either occupied ($f^{\dagger}_{i}f_{i}=1$) or empty
($f^{\dagger}_{i}f_{i}=0$), every term in the sum vanishes, and
$\Delta=0$ no matter what configuration of $f$\/-electrons we choose.

However, we argue that there are degenerate states in the $f$\/-occupation
representation, linear combinations of which are ground states with finite
polarization. It is straightforward to demonstrate this property for the exact
solutions of a ring of four sites. We construct a ring of four sites with
$d$\/- and $f$\/-states on alternate sites. The Hamiltonian, including all
the FK terms, has inversion symmetry with respect to the $d$\/- and
$f$\/-sites. There are three eigenstates which are even under inversion
with respect to a $d$\/-site, and three which are odd. The parameters are
varied so that a degeneracy occurs between two states of odd parity and
one of even parity. The ground state in the limit of a vanishing electric
field is a mixture of even and odd states with a finite polarization.

In Ref.~\onlinecite{Liu} an electronic-polaron solution of the FK model
is presented that may explain the anomalous properties of
heavy-electron materials such as ${\rm UPt}_{3}$, ${\rm UBe}_{13}$,
and ${\rm CeCu}_{2}{\rm Si}_{2}$. It is proposed that the Coulomb
interaction causes the $f$\/-electrons in these materials to propagate
like polarons, with a screening cloud made up of $d$\/-electrons.
The model is shown to account for a large number of thermodynamic and
transport properties of heavy-electron materials.

Here we show that the electronic polaron has a nonzero built-in
coherence
\begin{equation}
\Delta = \frac{U}{N} \sum_{\bf k} G_{d\/f}({\bf k},\tau=0^{-}) ,
\end{equation}
where $G_{d\/f}({\bf k},\tau)=-\langle T_{\tau} d_{\bf k}(\tau)
f^{\dagger}_{\bf k}(0) \rangle$ is a mixed Green's function with
Fourier transform $G_{d\/f}({\bf k},i\omega_{n}) = -V G_{f\/f}
({\bf k},i\omega_{n}) G_{d\/d}({\bf k},i\omega_{n})$, where $G_{d\/d}$
and $G_{f\/f}$ are the $d$\/- and $f$\/-electron propagators, and
$\omega_{n}$ are the Matsubara frequencies. Expressing the Green's
functions in the Lehmann representation, and summing over the Matsubara
frequencies, we find (at $T=0$)
\begin{equation}
\Delta = V U A \pi {\rm csc}(\pi \alpha)\,
\int_{\mu_{0}}^{\infty} d\omega\, \omega^{-\alpha} N_{+}(\omega) .
\end{equation}
Here $A$ and $\alpha$ are the forefactor and the singularity index
of the $f$\/-electron spectral function (see Ref.~\onlinecite{Liu}),
$N_{+}(\omega)$ is the density of $d$\/-electron states in the upper
branch, and $\mu_{0}$ is the Fermi level at $T=0$. Using Eq.~(8) of
Ref.~\onlinecite{Liu} for $N_{+}(\omega)$ we obtain
\begin{equation}
\Delta = \frac{U}{4V}\left(\frac{\eta}{\mu_{0}}\right)^{2\alpha} W ,
\end{equation}
where $W$ is the bandwidth and $\eta$ the upper-branch threshold.
Taking the parameter values $\alpha = 0.8$, $\mu_{0} = 1.3\,\eta$,
and $U = 20\,V$ from Ref.~\onlinecite{Liu} we get $\Delta = 3.28\,W$.

\section{Summary and conclusion}
\label{sec:SUMMARY}

In this paper we have investigated the linear and nonlinear optical
characteristics of the Falicov-Kimball model.
The SCMF solution of the periodic model results
 in the Bose-Einstein condensation of $d$\/-$f$ excitons. We found that
the pairing of $d$\/-states of even parity with $f$\/-states of odd
parity breaks the inversion symmetry of the underlying crystal,
leading to electronic ferroelectricity. The valence transition is
accompanied by a divergence of the static dielectric
constant at the critical value of the $f$\/-level energy.
The existence of electronic ferroelectricity in a given mixed-valent
compound predicates on the dominance of the $d$\/-$f$ Coulomb
interaction over the hybridization.

We have calculated both the linear and the second-harmonic susceptibilities
of a model mixed-valent system within the SCMF approximation. The absorption
spectrum of the mixed-valent system, when placed in an additional static
electric field, consists of a peak at the ferroelectric
resonance frequency and a continuum above the energy gap. The ferroelectric
resonance frequency is proportional to the square root of the effective
bias field, which depends, in addition to the applied static field,
 on the sample shape, domain structure, and crystal anisotropy. The continuum
absorption has a threshold singularity when the $f$\/-level lies within the
conduction band. The second-harmonic susceptibility is directly proportional to
the amount of built-in coherence $\Delta$. The final-state Coulomb interaction
enhances the second-harmonic conversion efficiency at $\omega=\Delta$ and
$\omega=2\Delta$. As experimental tests of the electronic ferroelectricity
 in mixed-valent compounds we proposed measurements of the static dielectric
constant, the microwave absorption spectrum, and the second-harmonic
susceptibility. The measured far-infrared absorption spectrum of SmB$_{6}$ was
found to be consistent with the model calculation.

We have also discussed the possibility of electronic ferroelectricity,
ferroelectric resonance, and second-harmonic generation for two other
theoretical ground states of the Falicov-Kimball model.
The electronic-polaron state does have a built-in coherence
comparable to the SCMF solution.
A ground state with $f$\/-occupation as a good quantum number has no
built-in coherence between $d$\/-electrons and $f$\/-holes. Such a
state on its own is not ferroelectric, does not exhibit ferroelectric
resonance, and cannot sustain second-harmonic generation.  We have
argued that degenerate ground states in the  $f$\/-occupation
representation can lead to ferroelectric ground states. The explicit
numerical demonstration of such ground states is left for the
future.

\section*{Acknowledgements}
We wish to thank M. B. Maple, S. H. Liu, D. P. Arovas, and S. R. Renn
for helpful discussions. This work was supported in part by NSF
Grant No.\ DMR 94-21966 and in part by the Deutsche Forschungsgemeinschaft
(DFG).

\section*{Appendix}
In this appendix we give the explicit expressions for the functions
$A(\omega)$ and $B(\omega)$ defined in Eqs.~(\ref{eq:FUNCA}) and
(\ref{eq:FUNCB}). Replacing $\frac{1}{N} \sum_{\bf k}$ by $\int
d\epsilon\rho(\epsilon)$, with $\rho(\epsilon)=\theta(W-|\epsilon|)/(2W)$,
we obtain
\begin{eqnarray}
A(\omega) & = & \frac{U}{2W} \int_{-W-E_{f}}^{W-E_{f}} d\epsilon
\frac{1}{\sqrt{\epsilon^{2}+M^{2}}(\omega^{2}-\epsilon^{2}-M^{2})} ,\\
B(\omega) & = & \frac{U}{2W} \int_{-W-E_{f}}^{W-E_{f}} d\epsilon
\frac{\epsilon}{\sqrt{\epsilon^{2}+M^{2}}(\omega^{2}-\epsilon^{2}-M^{2})}.
\end{eqnarray}
The integrals can be performed by elementary methods. We find
\begin{eqnarray}
\label{eq:EXPRA}
A(\omega) & = &  \frac{U}{4W^{2}\omega\sqrt{\omega^{2}-M^{2}}} \left[ \ln
\left(\frac{\sqrt{(E_{f}+W)^{2}+M^{2}}\sqrt{\omega^{2}-M^{2}}+(E_{f}+W)\omega}
           {\sqrt{(E_{f}+W)^{2}+M^{2}}\sqrt{\omega^{2}-M^{2}}-(E_{f}+W)\omega}
     \right) \right. \nonumber \\  & & - \left. \ln
\left(\frac{\sqrt{(E_{f}-W)^{2}+M^{2}}\sqrt{\omega^{2}-M^{2}}+(E_{f}-W)\omega}
           {\sqrt{(E_{f}-W)^{2}+M^{2}}\sqrt{\omega^{2}-M^{2}}-(E_{f}-W)\omega}
      \right) \right] ,
\end{eqnarray}
and
\begin{equation}
\label{eq:EXPRB}
B(\omega) = \frac{U}{4W^{2}\omega} \left[
\ln \left( \frac{\sqrt{(E_{f}+W)^{2}+M^{2}} + \omega}
                {\sqrt{(E_{f}+W)^{2}+M^{2}} - \omega} \right) -
\ln \left( \frac{\sqrt{(E_{f}-W)^{2}+M^{2}} + \omega}
                {\sqrt{(E_{f}-W)^{2}+M^{2}} - \omega} \right) \right] .
\end{equation}
The frequency $\omega$ is understood to have a positive imaginary
part $\delta$.  The function $A(\omega)$ is even in $E_{f}$, while
$B(\omega)$ is odd in $E_{f}$. This means $B(\omega)=0$ at half-filling.

The static susceptibility given by Eq.~(\ref{eq:CHI_MV}) involves the
functions $A(\omega)$ and $B(\omega)$ evaluated at $\omega=0$. Taking
the limit $\omega \rightarrow 0$ in Eqs.~(\ref{eq:EXPRA}) and
(\ref{eq:EXPRB}) we obtain
\begin{eqnarray}
\label{eq:A0}
A(0) & = & \frac{U}{2W^{2}M}
\left( \frac{W+E_{f}}{\sqrt{(W+E_{f})^{2}+M^{2}}} -
       \frac{W-E_{f}}{\sqrt{(W-E_{f})^{2}+M^{2}}} \right) , \\
\label{eq:B0}
B(0) & = & \frac{U}{2W^{2}}
\left( \frac{1}{\sqrt{(W-E_{f})^{2}+M^{2}}} -
       \frac{1}{\sqrt{(W+E_{f})^{2}+M^{2}}} \right) .
\end{eqnarray}

\begin{figure}
\caption{\label{fig:DELTA}
$f$\/-level occupancy $n_{f}$ and gap parameter $\Delta$ of the model
system as a function of the $f$\/-level energy $E_{f}$. The Coulomb
repulsion is $U = 3.0 W$.}
\end{figure}

\begin{figure}
\caption{\label{fig:EPS}
Static dielectric constant $\epsilon_{zz}$ of the model
system as a function of the $f$\/-level energy $E_{f}$. The inset
shows the built-in polarization as a function of $E_{f}$. The Coulomb
repulsion is $U = 15$ meV and the electric-dipole matrix element is
$\mu_{z} = 3.7\,10^{-29}$ Cm. The parameter values were obtained by
fitting the absorption spectrum of the model system to experimental
data on ${\rm SmB}_{6}$ (see Fig.~\protect\ref{fig:SMB}).}
\end{figure}

\begin{figure}
\caption{\label{fig:CHIONE}
Absorption spectrum ${\rm Im}\chi^{(1)}_{zz}(\omega)$ of the model
system as a function of the photon energy $\omega$, for various
values of the $f$\/-level energy $E_{f}$. The Coulomb repulsion is
$U = 3.0 W$. The solid lines show the continuum absorption spectrum
in zero bias field. The dash-dotted line shows both the ferroelectric
resonance peak and the continuum absorption spectrum in a bias field
of $E_{z} = 0.01 \mu_{z}/W$.}
\end{figure}

\begin{figure}
\caption{\label{fig:CHITWO}
Amplitude $|\chi^{(2)}_{zzz}(2\omega,\omega,\omega)|$ of the second-harmonic
susceptibility as a function of the photon energy $\omega$, for various
values of the $f$\/-level energy $E_{f}$. The dash-dotted line shows the
phase of $\chi^{(2)}_{zzz}(2\omega,\omega,\omega)$ for $E_{f}=-1.0 W$.
The Coulomb repulsion is $U=3.0 W$. The amplitude is given in units of
$N \mu_{z}^{3}/(2\Omega W^{2})$. For the parameter values given for the
solid line in Fig.~\protect\ref{fig:SMB}, $N \mu_{z}^{3}/(2\Omega W^{2})=
82\,{\rm nm V}^{-1}$.}
\end{figure}

\begin{figure}
\caption{\label{fig:SMB}
Comparison of the mean-field (solid line) and single-quasiparticle
(dash-dotted line) results for the infrared absorption spectrum
of ${\rm SmB}_{6}$ to experimental data taken from
 Ref.~\protect\onlinecite{Wachter} (diamonds).
The $f$\/-level energy is $E_{f}=0$, the $d$\/-bandwidth is
$W=40$ meV, and the Coulomb repulsion is $U=15$ meV. The
electric-dipole matrix element is $\mu_{z}=3.7\,10^{-29}$ Cm for
the solid line, and $\mu_{z}=4.2\,10^{-29}$ Cm for the dash-dotted
line.}
\end{figure}

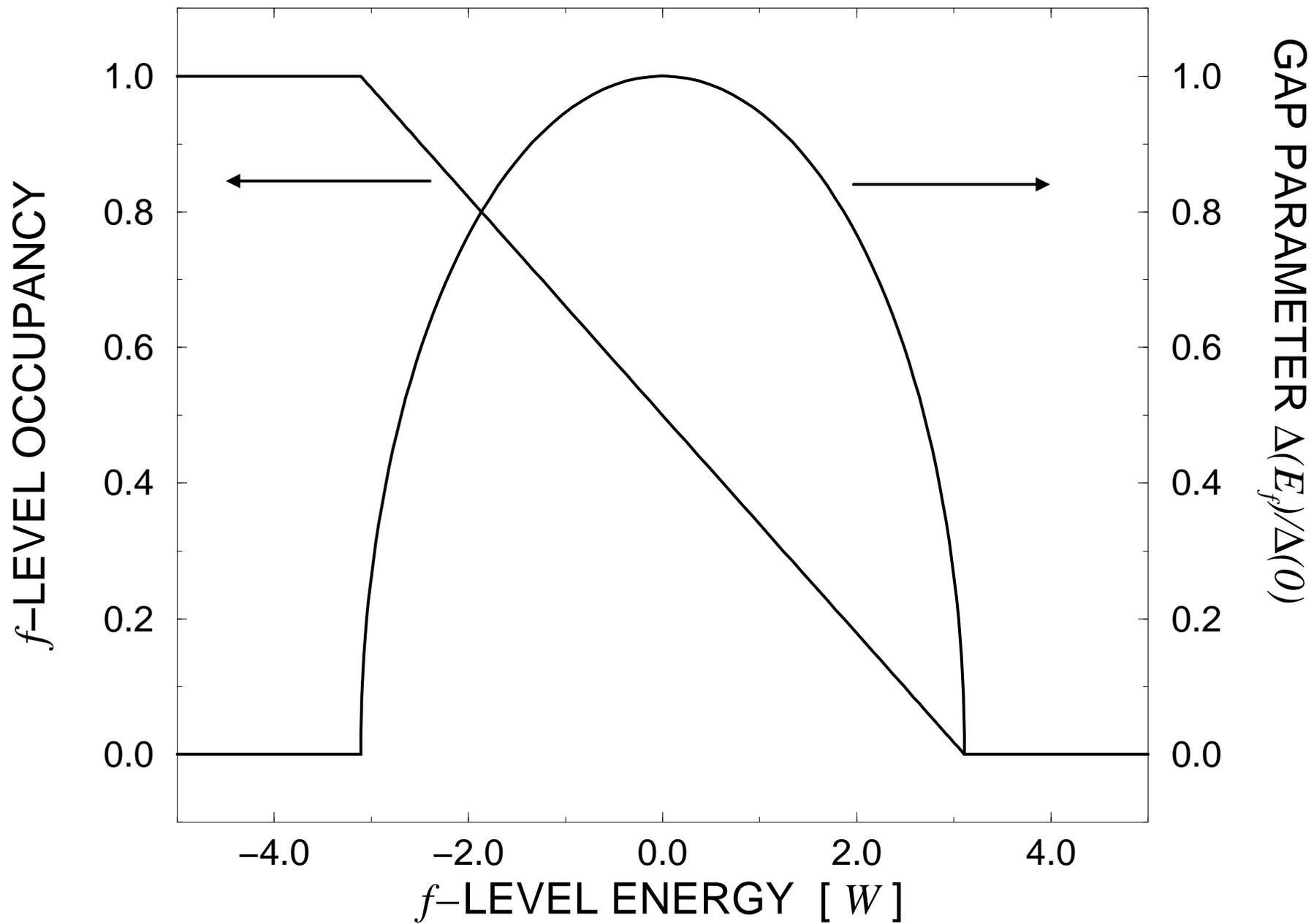
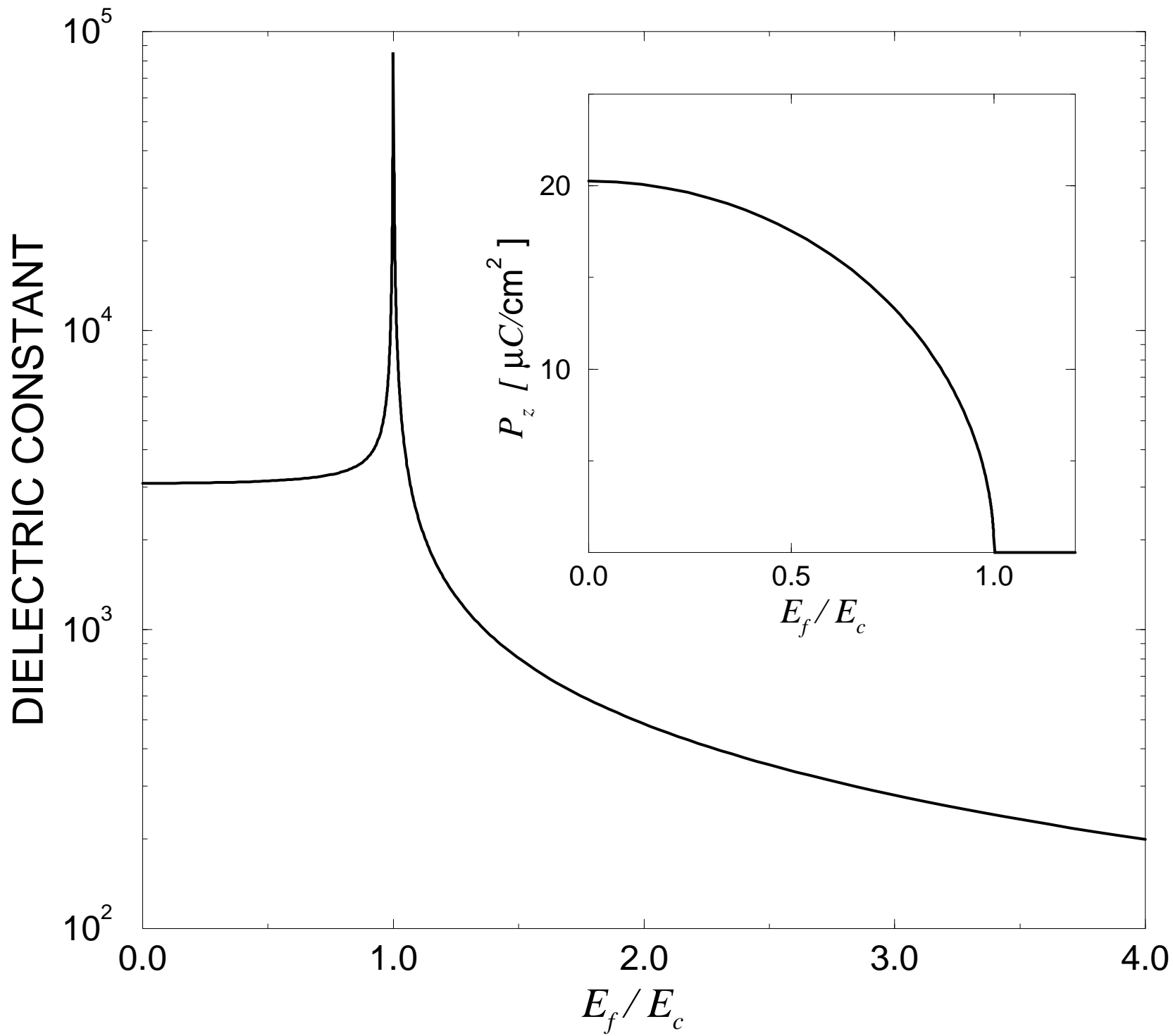
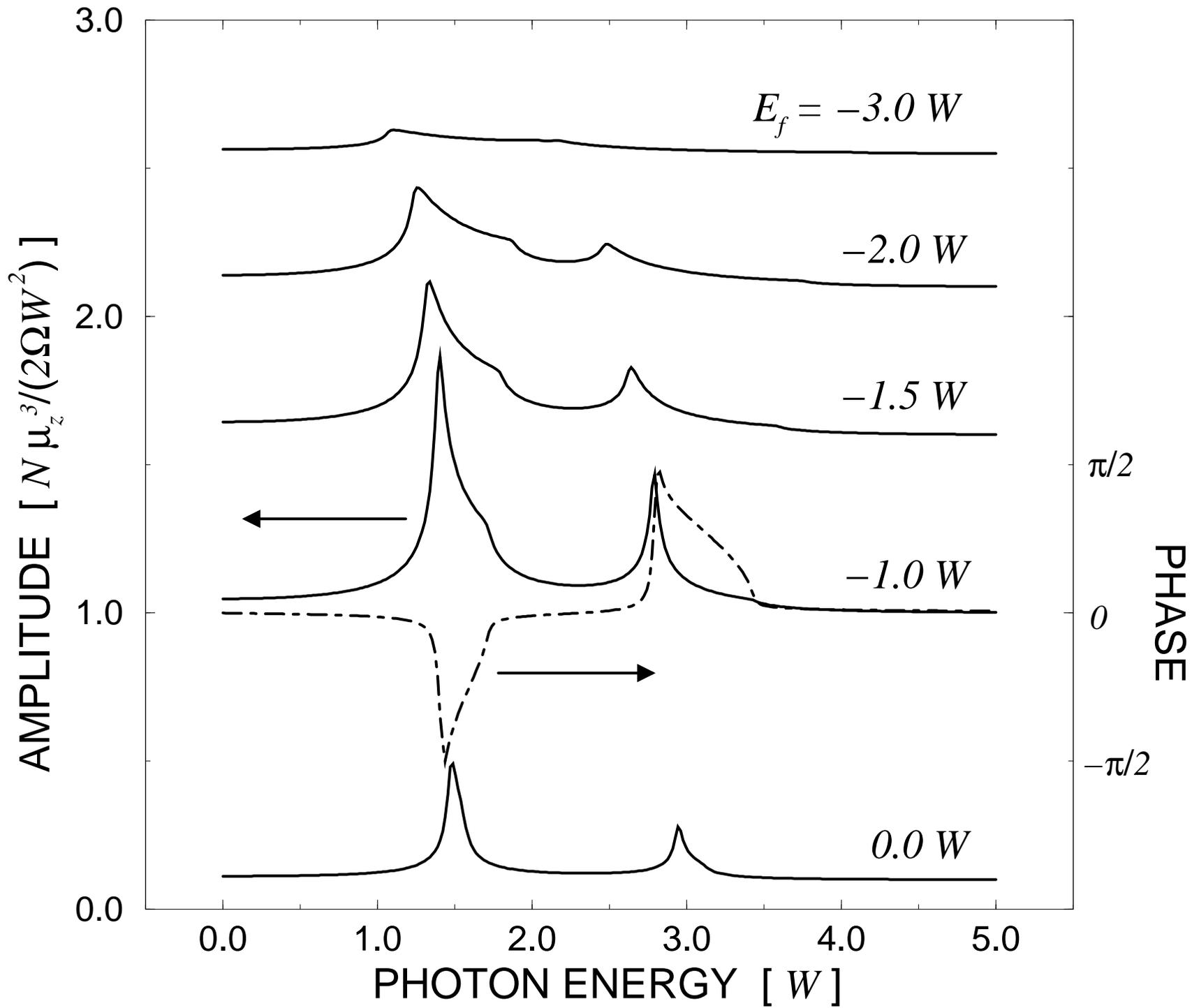
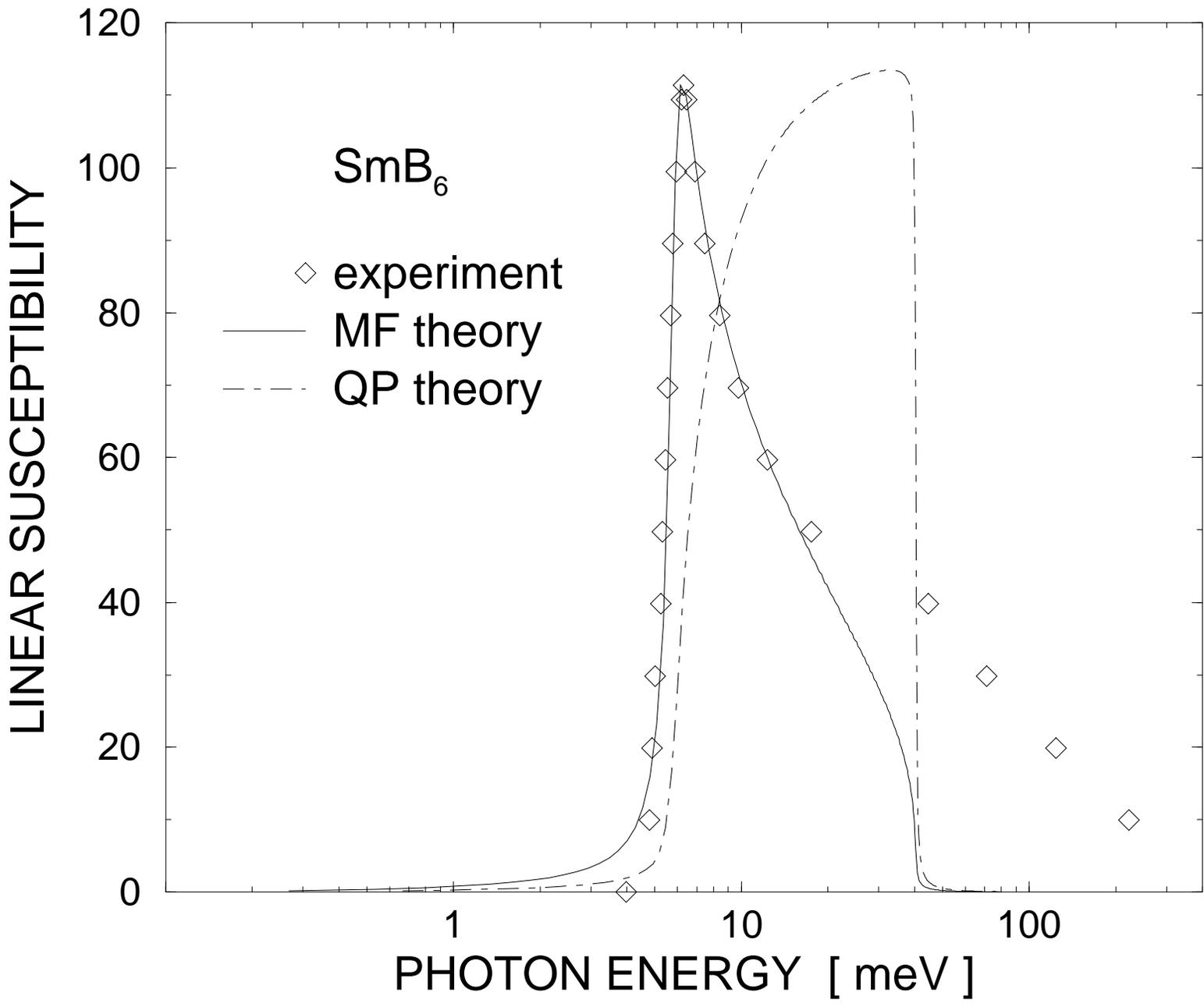
\end{document}